\documentclass[a4paper]{aa}
\usepackage{graphics,txfonts,amsfonts}
\pdfoutput=1

\usepackage{natbib}
\bibpunct[, ]{(}{)}{;}{a}{}{,}

\usepackage[usenames,dvipsnames]{xcolor}

\newcommand{\teff}{$T_{\rm eff}$}

\newcommand{\vs}{$v_{\rm e}\sin i$}
\newcommand{\bz}{$\langle B_{\rm z}\rangle$}
\newcommand{\kms}{km\,s$^{-1}$}

\newcommand{\peg}{II\,Peg}
\newcommand{\inv}{{\sc Invers13}}
\newcommand{\invb}{{\sc Invers10}}

\newcommand{\figps}[1]{\resizebox{\hsize}{!}{\rotatebox{0}{\includegraphics{#1}}}}

\newcommand{\fifps}[2]{\centering\resizebox{#1}{!}{\includegraphics{#2}}}
\newcommand{\firrps}[2]{\centering\resizebox{#1}{!}{\rotatebox{90}{\includegraphics{#2}}}}

\newcommand{\twofig}[2]{%
\resizebox{17cm}{!}{\rotatebox{90}{\includegraphics{#1}}}\\\medskip
\resizebox{17cm}{!}{\rotatebox{90}{\includegraphics{#2}}}
}

\newcommand{\beq}{\begin{equation}}
\newcommand{\eeq}{\end{equation}}

\begin{document}

\title{Magnetic field topology of the RS CVn star II Pegasi%
\thanks{Based on observations obtained with the Nordic Optical Telescope, operated on the island of La Palma jointly by Denmark, Finland, Iceland, Norway, and Sweden, at the Spanish Observatorio del Roque de los Muchachos of the Instituto Astrofisica de Canarias.}}

\author{O. Kochukhov\inst{1}
   \and M.J. Mantere\inst{2}
   \and T. Hackman\inst{2,3}
   \and I. Ilyin\inst{4}
   }


\institute{Department of Physics and Astronomy, Uppsala University, Box 516, Uppsala SE-751 20, Sweden\\ \email{oleg.kochukhov@fysast.uu.se}
      \and Department of Physics, Gustaf H\"allstr\"omin katu 2a, University of Helsinki, Box 64, Helsinki FI-00014, Finland
      \and Finnish Centre for Astronomy with ESO (FINCA), University of Turku, V\"ais\"al\"antie 20, Piikki\"o FI-21500, Finland
      \and Leibniz Institute for Astrophysics Potsdam (AIP), An der Sternwarte 16, Potsdam D-14482, Germany}

\date{Received 24 September 2012 / Accepted 10 December 2012}

\abstract%
{
The dynamo processes in cool active stars generate complex magnetic fields responsible for prominent surface stellar activity and variability at different time scales. For a small number of cool stars magnetic field topologies were reconstructed from the time series of spectropolarimetric observations using the Zeeman Doppler imaging (ZDI) method, often yielding surprising and controversial results.
}
{
In this study we follow a long-term evolution of the magnetic field topology of the RS~CVn binary star II\,Peg using a more self-consistent and physically more meaningful modelling approach compared to previous ZDI studies.
}
{
We collected high-resolution circular polarisation observations of II\,Peg using the SOFIN spectropolarimeter at the Nordic Optical Telescope. These data cover 12 epochs spread over 7 years, comprising one of the most comprehensive spectropolarimetric data sets acquired for a cool active star. A multi-line diagnostic technique in combination with a new ZDI code is applied to interpret these observations.
}
{
We have succeeded in detecting clear magnetic field signatures in average Stokes $V$ profiles for all 12 data sets. These profiles typically have complex shapes and amplitudes of $\sim$$10^{-3}$ of the unpolarised continuum, corresponding to mean longitudinal fields of 50--100~G. Magnetic inversions using these data reveals evolving magnetic fields with typical local strengths of 0.5--1.0~kG and complex topologies. Despite using a self-consistent magnetic and temperature mapping technique, we do not find a clear correlation between magnetic and temperature features in the ZDI maps. Neither do we confirm the presence of persistent azimuthal field rings found in other RS\,CVn stars. Reconstruction of the magnetic field topology of II\,Peg reveals significant evolution of both the surface magnetic field structure and the extended magnetospheric field geometry on the time scale covered by our observations. From 2004 to 2010 the total field energy drastically declined and the field became less axisymmetric. This also coincided with the transition from predominantly poloidal to mainly toroidal field topology. 
}
{
A qualitative comparison of the ZDI maps of II\,Peg with the prediction of dynamo theory suggests that the magnetic field in this star is produced mainly by the turbulent $\alpha^2$ dynamo rather than the solar $\alpha\Omega$ dynamo. Our results do not show a clear active longitude system, nor is there an evidence of the presence of an azimuthal dynamo wave.
}

\maketitle

\section{Introduction}
\label{intro}

Direct, spatially-resolved observations of the solar disk demonstrate the ubiquitous presence of magnetic fields and their key role in driving time-dependent, energetic processes. It is believed that magnetic fields are generated in the solar interior, due to a complex interplay between the non-uniformities in the internal rotation profile, large-scale flows, and vigorous turbulence due to convective motions \citep{ossendrijver:2003}. Although the rotation profile, with two prominent regions of high radial shear near the bottom and top of the solar convection zone, is known rather reliably from helioseismic observations \citep[e.g.][]{thompson:2003}, measurements of the potentially very important meridional flow pattern, especially deeper in the convection zone, are highly uncertain \citep{gough:2010}. A variety of models of the solar dynamo exists \citep[e.g.][]{dikpati:1999,kapyla:2006,kitchatinov:2012}, but none of them is capable of fully reproducing the observed characteristics of the solar magnetic cycle. Direct numerical simulations aiming at including all the relevant physics in a single model, with the expectation being that all the observable features from the flow patterns up to the magnetic cycle would emerge self-consistently, have so far been only marginally successful \citep{ghizaru:2010,racine:2011,kapyla:2012a}. Therefore, it is fair to say that the solar cycle remains a theoretical challenge, with some key information on the large-scale flow patterns yet to be supplied by observations.

Compared to the Sun, many cool stars exhibit enhanced activity levels, suggesting the presence of stronger surface magnetic fields produced by more efficient dynamos. Observations of such active stars offer possibilities to constrain theoretical models by examining the efficiency of dynamo mechanisms as a function of stellar mass, age, and rotation rate. However, direct magnetic field analysis of cool active stars is impeded by the weakness of polarisation signatures in the spatially-unresolved stellar spectra. Furthermore, dynamos in cool stars usually give rise to topologically complex magnetic fields which cannot be meaningfully characterised using observational techniques sensitive only to the global magnetic field component \citep{borra:1980,vogt:1980}.

A major breakthrough in studying the magnetism of cool stars was achieved by combining multi-line polarisation diagnostic techniques \citep{donati:1997} applied to high-resolution spectropolarimetric observations with the inversion procedure of Zeeman Doppler imaging \citep[ZDI,][]{brown:1991,donati:1997a}. This approach, capable of providing spatially-resolved information about stellar magnetic field topologies, was applied to a number of late-type stars \citep[e.g.][]{donati:2003,petit:2004a}, but only three objects -- AB\,Dor, LQ\,Hya, and HR\,1099 -- were systematically followed with multiple magnetic maps on the time scales of five to ten years. 

The results of ZDI mapping are often spectacular in their ability to resolve fine details of stellar magnetic field geometries, but also controversial in several aspects. First, analysis of stellar spectra is limited to only circular polarisation because, being roughly ten times weaker, the linear polarisation signatures are not readily detectable at the moderate signal-to-noise (S/N) ratios used for the circular polarisation monitoring \citep{kochukhov:2011}. Supplying inversion algorithms with such incomplete Stokes vector data is known to lead to spurious features in the reconstructed magnetic maps \citep{donati:1997a,kochukhov:2002c}. Second, interpretation of the mean-line shapes constructed by averaging thousands of lines is inevitably inferior to modelling individual spectral lines because of the difficulty in choosing appropriate mean-line parameters \citep{kochukhov:2010a}. Finally, and most importantly, nearly all magnetic field maps reconstructed from the Stokes $V$ observations of cool active stars were obtained separately and inconsistently from the mapping of brightness distributions from Stokes $I$, raising questions about the validity of magnetic maps and preventing direct analysis of the spatial relation between magnetic and temperature inhomogeneities.

These considerations suggest that substantial progress in understanding cool-star magnetism through ZDI requires continuing research in several directions. On the one hand, the sample of active stars studied with multiple-epoch magnetic images has to be expanded to justify the far-reaching conclusions previously inferred from the analysis of a few objects. On the other hand, ZDI studies must re-examine key methodological limitations of this technique and strive to employ physically realistic modelling approaches whenever possible. This paper addresses both of these aspects in a detailed ZDI analysis of the RS~CVn binary \peg.

The source of the most powerful stellar flares ever observed \citep{osten:2007} and the brightest X-ray object within 50~pc \citep{makarov:2003}, \peg\ (HD\,224085, HIP\,117915) is one of the most prominent active cool stars in the solar neighbourhood. This star is an RS~CVn-type, single-line spectroscopic binary (SB1) with an orbital period of $\approx$\,6.72 days, consisting of a K2IV primary and a low-mass (M0--M3V) secondary. The primary star exhibits copious manifestations of the magnetically-driven surface activity, including a strong non-thermal emission in the UV and optical chromospheric lines and in the X-ray and radio wavelength regions. It also shows powerful flares as well as regular photometric and spectroscopic variations due to evolving cool spots. \citet{berdyugina:1998} studied the orbital motion of the massive component and provided a comprehensive summary of the physical properties of \peg. Many studies examined photometric variations of this star \citep[e.g.][and references therein]{siwak:2010,roettenbacher:2011}, aiming to explore long-term activity cycles and, in particular, to investigate the role of active longitudes \citep{berdyugina:1999,rodono:2000}.

Other studies targeted \peg\ with high-resolution spectroscopic observations with the goal to constrain the spot temperatures using TiO absorption bands \citep{oneal:1998,berdyugina:1998} and to analyse configurations of the surface temperature inhomogeneities with the Doppler imaging technique \citep{berdyugina:1998b,gu:2003,lindborg:2011,hackman:2012}. Early DI images corresponding to the period between 1994 and 2002 suggested persistent presence of a pair of active longitudes and showed major changes in the surface structure on a time-scale of less than a year \citep{berdyugina:1998b,berdyugina:1999,lindborg:2011}. More recent DI maps covering the years from 2004 to 2010 revealed the star entering a low-activity state characterised by a more random distribution of cool spots \citep{hackman:2012}.

The underlying cause of the remarkable surface activity of \peg\ -- the dynamo-generated magnetic field -- was first detected in this star by \citet{donati:1992} with four circular polarisation observations of a few magnetically sensitive lines. The presence of the Stokes $V$ signatures in spectral lines was subsequently confirmed by \citet{donati:1997} using a multi-line polarimetric analysis, but no systematic phase-resolved investigation of the magnetic field topology of \peg\ has ever been undertaken.

Since 2004 we have been monitoring the magnetic field in \peg\ using the SOFIN spectropolarimeter at the Nordic Optical Telescope. We have acquired a unique collection of high-resolution Stokes $I$ and $V$ spectra covering $\approx$\,5.5 years or almost 300 stellar rotations. In this paper we present a comprehensive analysis of these polarisation data, focusing on the self-consistent ZDI mapping of stellar magnetic field topology. The accompanying study by \citet{hackman:2012} obtained temperature maps from the same SOFIN Stokes $I$ spectra. Preliminary attempts to map the magnetic field geometry of \peg, using about 10\% of the spectropolarimetric data analysed here, were presented first by \citet{carroll:2007}, and then by \citet{carroll:2009a} and \citet{kochukhov:2009d}.

This paper is organised as follows. In Sect.~\ref{obs} we describe the acquisition and reduction of the spectropolarimetric observations of \peg. Detection of the magnetic signatures in spectral lines with the help of a multi-line analysis is presented in Sect.~\ref{lsd}. The methodology of the self-consistent ZDI and the choice of stellar parameters required for mapping is described in Sect.~\ref{zdi}. Results of the magnetic and temperature inversions of \peg\ are presented and analysed in Sect.~\ref{res}. The outcome of our investigation is discussed in the context of previous observational and theoretical studies in Sect.~\ref{disc}.

\section{Spectropolarimetric observations}
\label{obs}

The spectropolarimetric observations of \peg\ analysed here were carried out during the period from Jul. 2004 to Jan. 2010 using the SOFIN echelle spectrograph \citep{tuominen:1999} at the 2.56-m Nordic Optical Telescope. The spectrograph, which is mounted at the Cassegrain focus and is equipped with a $2048\times2048$ pixel CCD detector, was configured to use its second camera yielding resolving power of $R$\,$\approx$\,70000. For ZDI analysis we used 12 echelle orders, each covering 40--50~\AA\ in the wavelength region between 4600 and 6135~\AA. The data for \peg\ were obtained during 12 individual epochs, for which from 3 to 12 observations were recorded over the time span ranging from 3 to 14 nights. The spectra have typical S/N ratio of 200--300.

The circular polarisation observations were obtained with the Zeeman analyser, consisting of a calcite plate used as a beam splitter and an achromatic rotating quarter-wave plate. At least two exposures with the quarter-wave retarder angles separated by 90\degr\ is required to obtain the Stokes $V$ spectrum. Rotation of the quarter-wave plate has the effect of exchanging positions of the right- and left-hand circularly polarised beams on the detector. The beam exchange procedure \citep{semel:1993} facilitates an accurate polarisation analysis because possible instrumental artefacts change sign when the quarter-wave plate is rotated and then cancel out when all sub-exposures are combined. To accumulate sufficient signal-to-noise ratio, a sequence of 2 or 3 double exposures was obtained in this way. The length of individual sub-exposures varied between 15 and 25~min, depending on seeing and weather conditions. 

The data were reduced with the help of the {\sc 4A} software package \citep{ilyin:2000}. Specific details of the SOFIN polarimeter design and corresponding data reduction methods are given by \citet{Ilyin:2012}. The spectral processing included standard reduction steps, such as bias subtraction, flat field correction, removal of the scattered light, and optimal extraction of the spectra. Wavelength calibration used ThAr exposures obtained before and after each single exposure to account for environmental variations and gravitational bending of a Cassegrain mounted spectrograph by means of a global fit of the two ThAr wavelength solutions versus time.

Spectropolarimetric observations of strongly magnetic Ap stars are frequently performed with SOFIN in a configuration similar to the one used for \peg. These measurements agree closely with the results obtained for the same stars at other telescopes \citep[e.g.][]{ryabchikova:2007a,Ilyin:2012}. Repeated observations of a very slowly rotating Ap star $\gamma$~Equ reveals no systematic differences in the Stokes $V$ profiles during the entire period of our observations of \peg. This confirms the robustness and accuracy of the employed instrument calibration and data reduction methods.

The log of all 88 SOFIN Stokes $V$ observations of \peg\ is given in Table~\ref{tbl:obs}. The rotation of the primary component of \peg\ is synchronised with the orbital motion. Therefore, we calculated rotational phases using the orbital ephemeris from \citet{berdyugina:1998}
\beq
T = 2449582.9268 + 6.724333 \times E
\eeq
which refers to the time of orbital conjunction. The orbital solution from the same study was applied to correct the radial velocity variation caused by the orbital motion.

Our 12 sets of spectropolarimetric observations of \peg\ comprise 3 to 12 distinct rotational phases, whereas $\sim$\,10 evenly distributed phases are needed for an optimal ZDI reconstruction \citep{kochukhov:2002c}. Assuming that each observation provides a coverage of 10\% of the rotational period, we quantified the phase coverage of each data set as a fraction $f$ of the full rotational cycle (see Table~\ref{tbl:obs}). Our SOFIN observations yield from $f$\,=\,30\% (epoch 2005.0) to $f$\,=\,87\% (epoch 2009.7). The three epochs with the best phase-coverage are 2004.6 (75\%), 2007.6 (81\%), and 2009.7 (87\%). 

The quality of ZDI mapping is reduced for the data sets with an insufficient phase-coverage. Yet, as demonstrated by numerical experiments \citep{donati:1997a,kochukhov:2002c} and ZDI studies based on observations with poor phase-coverage \citep{donati:1999b,hussain:2009}, it is still possible to extract useful information about the stellar surface features from just a few spectra and from observations covering only half of the stellar rotation cycle. Aiming to maintain consistency in our long-term ZDI analysis of \peg, we performed magnetic inversions for all 12 epochs. At the same time, we kept in mind the difference in data quality during the assessment of the inversion results.

\onllongtab{
\begin{small}
\begin{longtable}{cccccrrr}
\caption{The journal of spectropolarimetric observations of \peg. \label{tbl:obs}}\\
\hline\hline
Epoch/ & UT Date & HJD    & Rotational & S/N & S/N~~~ & LSD $V_{\rm max}$ & \bz~~~~ \\
Phase coverage &  & $-2400000$ & phase & ($V$) & (LSD $V$) & $\times10^4$~~ & (G)~~~~ \\
\hline
\endfirsthead
\caption{Continued.}\\
\hline\hline
Epoch/ & UT Date & HJD    & Rotational & S/N & S/N~~~ & LSD $V_{\rm max}$ & \bz~~~~ \\
Phase coverage &  & $-2400000$ & phase & ($V$) & (LSD $V$) & $\times10^4$~~ & (G)~~~~ \\
\hline
\endhead
\hline
\endfoot
2004.6/75\% & 30/07/2004 & 53216.6176 & 0.379 & 203 &  7842 & 14.5 & $  34.4\pm 6.6$ \\
       & 31/07/2004 & 53217.6363 & 0.531 & 282 & 11941 & 12.4 & $ -11.7\pm 4.3$ \\
       & 01/08/2004 & 53218.6476 & 0.681 & 268 & 10987 &  6.3 & $  12.2\pm 4.6$ \\
       & 02/08/2004 & 53219.6440 & 0.829 & 267 & 10809 & 12.8 & $ 145.8\pm 4.7$ \\
       & 03/08/2004 & 53220.6525 & 0.979 & 216 &  8909 &  8.2 & $ -49.8\pm 5.7$ \\
       & 05/08/2004 & 53222.6122 & 0.271 & 245 &  9894 & 15.2 & $  54.6\pm 5.2$ \\
       & 10/08/2004 & 53227.7210 & 0.031 & 290 & 11675 &  7.9 & $ 102.7\pm 4.4$ \\
       & 11/08/2004 & 53228.6151 & 0.164 & 299 & 12054 &  9.1 & $  55.2\pm 4.2$ \\[3pt]
2005.0/30\% & 30/12/2004 & 53370.3948 & 0.248 & 255 &  9896 & 15.7 & $  19.6\pm 5.1$ \\
       & 31/12/2004 & 53371.3766 & 0.394 & 208 &  8620 & 12.1 & $  85.2\pm 6.0$ \\
       & 01/01/2005 & 53372.3825 & 0.544 & 111 &  4511 & 10.8 & $ -12.1\pm11.2$ \\[3pt]
2005.6/56\% & 16/07/2005 & 53567.7037 & 0.591 & 228 &  8855 &  5.7 & $   8.8\pm 5.7$ \\
       & 17/07/2005 & 53568.7131 & 0.741 & 222 &  8567 &  9.4 & $  74.4\pm 5.9$ \\
       & 18/07/2005 & 53569.7274 & 0.892 & 224 &  8836 & 14.0 & $ 241.6\pm 5.7$ \\
       & 19/07/2005 & 53570.6443 & 0.028 & 247 &  8876 & 13.1 & $  56.5\pm 5.6$ \\
       & 20/07/2005 & 53571.6614 & 0.179 & 254 &  9979 & 13.1 & $ -34.8\pm 5.1$ \\
       & 23/07/2005 & 53574.6753 & 0.627 & 235 &  9528 &  6.7 & $  66.8\pm 5.3$ \\
       & 24/07/2005 & 53575.6487 & 0.772 & 250 &  9714 &  8.1 & $ 134.6\pm 5.2$ \\[3pt]
2005.9/38\% & 11/11/2005 & 53685.5055 & 0.109 & 133 &  4970 & 12.2 & $  12.5\pm10.3$ \\
       & 11/11/2005 & 53686.4788 & 0.254 & 159 &  5627 & 12.2 & $  87.6\pm 9.2$ \\
       & 17/11/2005 & 53692.4838 & 0.147 & 291 & 10684 & 15.8 & $   9.0\pm 4.7$ \\
       & 19/11/2005 & 53693.5483 & 0.305 & 248 &  9538 &  9.4 & $  72.7\pm 5.3$ \\
       & 20/11/2005 & 53695.4708 & 0.591 & 246 &  9484 &  8.6 & $  77.8\pm 5.4$ \\[3pt]
2006.7/72\% & 30/08/2006 & 53978.5023 & 0.682 & 234 &  8975 & 10.3 & $  35.7\pm 5.6$ \\
       & 01/09/2006 & 53979.6544 & 0.853 & 267 & 10100 & 14.8 & $ 184.4\pm 5.0$ \\
       & 02/09/2006 & 53980.6863 & 0.007 & 265 &  9516 & 20.0 & $ 102.5\pm 5.1$ \\
       & 05/09/2006 & 53983.6716 & 0.451 & 171 &  6323 &  6.1 & $  45.4\pm 8.0$ \\
       & 06/09/2006 & 53984.7310 & 0.608 & 233 &  8922 &  8.7 & $  52.1\pm 5.6$ \\
       & 07/09/2006 & 53985.6761 & 0.749 & 369 & 13796 & 12.1 & $  62.5\pm 3.7$ \\
       & 09/09/2006 & 53987.7055 & 0.051 & 368 & 13427 & 13.0 & $  83.1\pm 3.7$ \\
       & 11/09/2006 & 53989.7039 & 0.348 & 343 & 12866 &  6.7 & $  48.7\pm 4.0$ \\
       & 12/09/2006 & 53990.6496 & 0.488 & 438 & 16330 &  5.3 & $  89.6\pm 3.1$ \\
       & 13/09/2006 & 53991.6798 & 0.642 & 185 &  7181 &  8.2 & $   4.7\pm 7.1$ \\
       & 14/09/2006 & 53992.6805 & 0.790 & 232 &  8898 & 10.3 & $  86.6\pm 5.7$ \\[3pt]
2006.9/63\% & 01/12/2006 & 54071.4660 & 0.507 & 215 &  8461 &  8.2 & $  22.4\pm 5.9$ \\
       & 03/12/2006 & 54072.5264 & 0.665 & 162 &  5987 & 12.2 & $ -45.0\pm 8.2$ \\
       & 04/12/2006 & 54074.4128 & 0.945 & 199 &  7634 & 14.4 & $  64.0\pm 6.5$ \\
       & 05/12/2006 & 54075.4717 & 0.103 & 172 &  6316 & 14.3 & $ -10.8\pm 7.8$ \\
       & 06/12/2006 & 54076.4425 & 0.247 & 203 &  7614 & 11.1 & $ -63.6\pm 6.5$ \\
       & 07/12/2006 & 54077.4278 & 0.394 & 166 &  6691 & 10.8 & $  17.3\pm 7.5$ \\
       & 08/12/2006 & 54078.4148 & 0.540 & 148 &  5822 &  9.2 & $ -65.7\pm 8.6$ \\[3pt]
2007.6/81\% & 19/07/2007 & 54300.6961 & 0.597 & 253 &  9764 &  5.4 & $ -72.7\pm 5.3$ \\
       & 20/07/2007 & 54301.7103 & 0.747 & 290 & 10955 & 12.4 & $ -74.0\pm 4.7$ \\
       & 21/07/2007 & 54302.6741 & 0.891 & 299 & 11346 & 13.8 & $  33.8\pm 4.5$ \\
       & 22/07/2007 & 54303.6297 & 0.033 & 291 & 11222 &  9.2 & $  19.5\pm 4.6$ \\
       & 23/07/2007 & 54304.6726 & 0.188 & 265 &  9659 & 10.4 & $  41.4\pm 5.3$ \\
       & 24/07/2007 & 54305.7046 & 0.341 & 325 & 12704 &  6.7 & $  59.0\pm 4.0$ \\
       & 25/07/2007 & 54306.6431 & 0.481 & 343 & 13228 &  6.4 & $ -18.6\pm 3.9$ \\
       & 26/07/2007 & 54307.7051 & 0.639 & 266 & 10224 &  6.5 & $-108.0\pm 5.0$ \\
       & 27/07/2007 & 54308.6486 & 0.779 & 300 & 11357 & 15.6 & $ -81.1\pm 4.5$ \\
       & 28/07/2007 & 54309.6977 & 0.935 & 242 &  9075 & 10.5 & $  32.3\pm 5.6$ \\[3pt]
2007.9/63\% & 22/11/2007 & 54427.4366 & 0.445 & 264 & 10021 &  7.2 & $ -47.7\pm 5.1$ \\
       & 25/11/2007 & 54430.4725 & 0.896 & 166 &  5998 &  7.6 & $  58.1\pm 8.1$ \\
       & 26/11/2007 & 54431.4755 & 0.045 & 260 &  9451 & 16.3 & $  67.2\pm 5.3$ \\
       & 27/11/2007 & 54432.4542 & 0.191 & 378 & 10369 &  9.2 & $  59.8\pm 5.0$ \\
       & 30/11/2007 & 54435.3169 & 0.617 & 128 &  4469 & 14.1 & $-130.2\pm11.1$ \\
       & 01/12/2007 & 54436.4705 & 0.788 & 326 & 12251 &  7.0 & $ -21.8\pm 4.1$ \\
       & 02/12/2007 & 54437.4485 & 0.934 & 321 & 11758 &  7.3 & $  73.5\pm 4.3$ \\[3pt]
2008.7/40\% & 08/09/2008 & 54717.6059 & 0.597 & 328 & 12065 &  8.6 & $ -45.4\pm 4.3$ \\
       & 11/09/2008 & 54720.5529 & 0.035 & 313 & 12043 &  2.1 & $  -1.7\pm 4.3$ \\
       & 13/09/2008 & 54722.6254 & 0.343 & 348 & 12863 & 12.0 & $  34.3\pm 4.0$ \\
       & 14/09/2008 & 54723.5442 & 0.480 & 330 & 12382 &  7.6 & $  26.5\pm 4.2$ \\[3pt]
2008.9/59\% & 08/12/2008 & 54809.3926 & 0.247 & 415 & 15398 &  7.0 & $  45.5\pm 3.4$ \\
       & 09/12/2008 & 54810.4185 & 0.399 & 114 &  3932 & 11.5 & $   8.6\pm12.2$ \\
       & 10/12/2008 & 54811.4780 & 0.557 & 242 &  8764 & 10.0 & $ -41.6\pm 5.8$ \\
       & 11/12/2008 & 54812.4383 & 0.700 & 283 & 10017 &  7.9 & $ -52.2\pm 5.0$ \\
       & 13/12/2008 & 54814.4361 & 0.997 & 368 & 13643 &  3.1 & $  34.2\pm 3.7$ \\
       & 14/12/2008 & 54815.4722 & 0.151 & 292 & 10561 &  3.6 & $  20.1\pm 4.8$ \\[3pt]
2009.7/87\% & 26/08/2009 & 55069.6762 & 0.954 & 328 & 12128 &  7.0 & $   1.4\pm 4.2$ \\
       & 27/08/2009 & 55070.6893 & 0.105 & 310 & 11060 &  8.4 & $ -72.2\pm 4.6$ \\
       & 28/08/2009 & 55071.7238 & 0.259 & 200 &  7003 & 19.8 & $ -35.4\pm 7.2$ \\
       & 29/08/2009 & 55072.6965 & 0.404 & 347 & 12219 & 13.7 & $  53.6\pm 4.2$ \\
       & 30/08/2009 & 55073.7070 & 0.554 & 202 &  7326 & 20.1 & $  67.1\pm 7.0$ \\
       & 01/09/2009 & 55075.5550 & 0.829 & 274 &  9898 & 13.8 & $  40.6\pm 5.1$ \\
       & 02/09/2009 & 55076.6844 & 0.997 & 393 & 15026 &  4.3 & $ -19.5\pm 3.4$ \\
       & 03/09/2009 & 55077.6378 & 0.138 & 443 & 16263 & 10.7 & $ -49.8\pm 3.2$ \\
       & 04/09/2009 & 55078.7420 & 0.303 & 359 & 13033 & 14.9 & $  -2.3\pm 3.9$ \\
       & 05/09/2009 & 55079.6788 & 0.442 & 372 & 13695 & 11.2 & $  36.4\pm 3.7$ \\
       & 06/09/2009 & 55080.6736 & 0.590 & 374 & 12305 & 19.7 & $  59.9\pm 4.2$ \\
       & 07/09/2009 & 55081.7424 & 0.749 & 177 &  5654 & 11.0 & $  30.5\pm 9.2$ \\[3pt]
2010.0/68\% & 27/12/2009 & 55193.3636 & 0.348 & 296 & 10102 & 10.9 & $   3.4\pm 5.0$ \\
       & 28/12/2009 & 55194.3791 & 0.499 & 287 &  9793 & 13.4 & $  55.5\pm 5.1$ \\
       & 29/12/2009 & 55195.3684 & 0.647 & 397 & 14431 &  7.3 & $  32.9\pm 3.6$ \\
       & 31/12/2009 & 55197.3984 & 0.948 & 397 & 14261 & 10.8 & $  -9.4\pm 3.6$ \\
       & 01/01/2010 & 55198.4116 & 0.099 & 338 & 12572 &  5.0 & $ -57.5\pm 4.1$ \\
       & 02/01/2010 & 55199.3720 & 0.242 & 283 & 10120 & 16.7 & $ -63.2\pm 5.0$ \\
       & 03/01/2010 & 55200.3627 & 0.389 & 230 &  8445 & 14.4 & $  -4.6\pm 5.9$ \\
       & 04/01/2010 & 55201.3844 & 0.541 & 235 &  8599 & 13.3 & $  50.8\pm 5.9$ \\
\end{longtable}
\end{small}
}

\section{Multi-line polarisation analysis}
\label{lsd}

\subsection{Least-squares deconvolution}

\begin{figure}[!th]
\figps{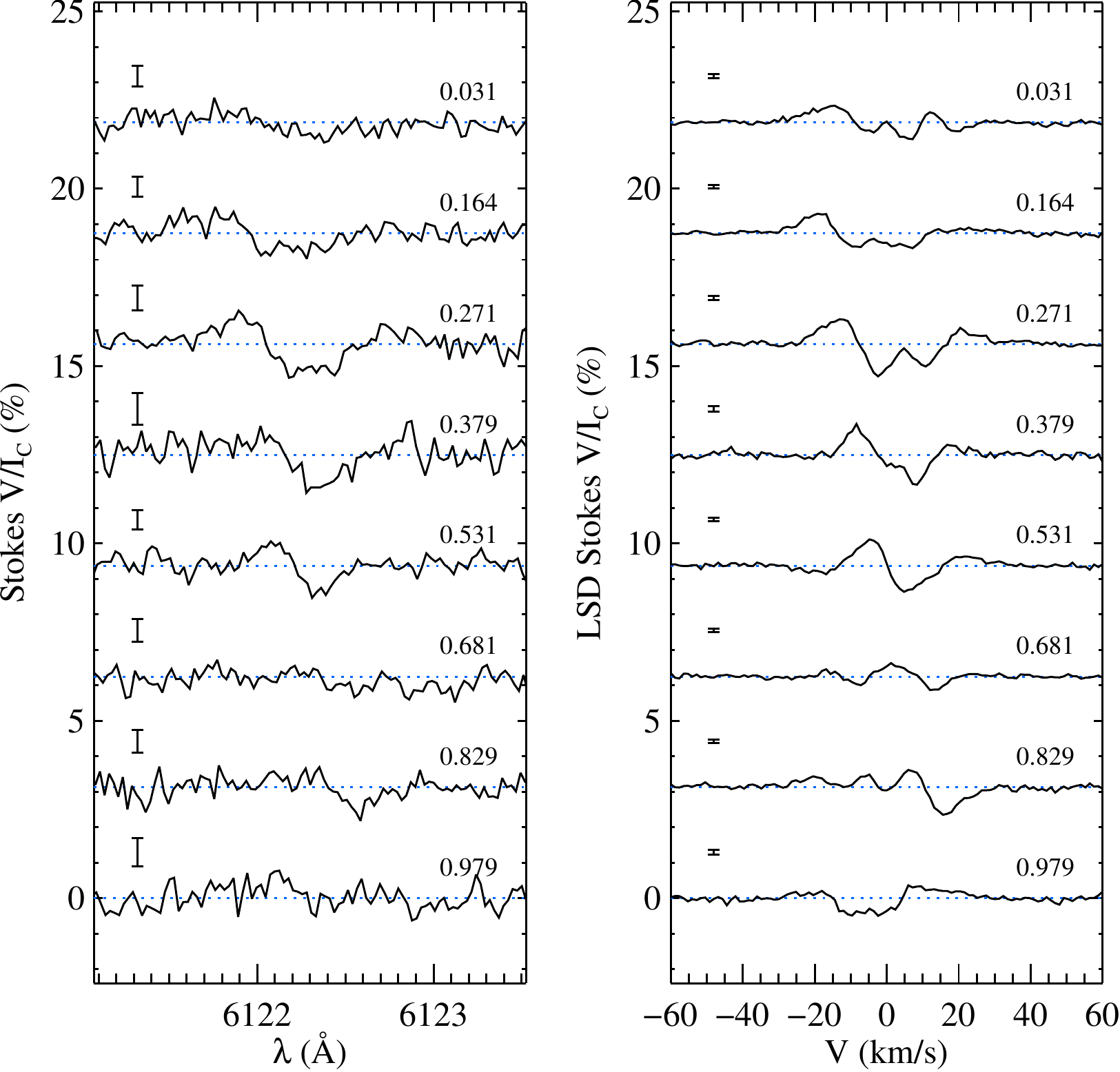}
\caption{Comparison of the Stokes $V$ profiles of the \ion{Ca}{i} 6122.2~\AA\ line (\textit{left panel}) with the scaled LSD $V$ profiles (\textit{right panel}) at epoch 2004.6. The spectra for different rotational phases are shifted vertically. The 1-$\sigma$ error bars are indicated on the left side of each panel.
}
\label{fig:lsd-sp}
\end{figure}

Even the most active late-type stars exhibit a relatively low amplitude of the circular polarisation signal in spectral lines, rarely exceeding 1\% of the Stokes $V$ continuum intensity. Given the moderate quality of the SOFIN spectropolarimetric observations of \peg, we were able to detect Stokes $V$ signatures at the 2--3$\sigma$ confidence level in only a few of the strongest spectral lines (Fig.~\ref{fig:lsd-sp}). This makes it challenging to model the magnetic field of \peg\ using a direct analysis of individual line profiles. A widely-used approach to overcome this difficulty is to employ a multi-line technique, combining information from hundreds or thousands of individual metal lines. 

In this study of \peg\ we applied the least-squares deconvolution (LSD) technique \citep{donati:1997} using the code and methodology described by \citet{kochukhov:2010a}. The LSD technique extracts information from all available lines by assuming that the Stokes $I$ and $V$ spectra can be represented by a superposition of corresponding scaled mean profiles. The scaling factors, established under the weak-line and weak-field approximations, are equal to the central line depth $d$ for Stokes $I$ and to the product of the line depth, laboratory wavelength of the transition $\lambda$, and its effective Land\'e factor $z$ for Stokes $V$, respectively. A linear superposition of scaled profiles is mathematically equivalent to a convolution of the average profile with a line mask. This simple stellar spectrum model can be inverted, obtaining a high-quality mean profile for a given line mask and observational data. \citet{kochukhov:2010a} showed that, for the magnetic fields below $\approx$\,2~kG, the LSD Stokes $V$ profiles derived in this way can be interpreted as a real Zeeman triplet line with an average Land\'e factor.

The atomic line data required for the application of LSD to observations of \peg\ were obtained from the {\sc vald} database \citep{kupka:1999}. We extracted a total of 1580 spectral lines with the central depth larger than 10\% of the continuum. The line intensities were calculated for a {\sc marcs} model atmosphere \citep{gustafsson:2008} with the effective temperature $T_{\rm eff}$\,=\,$4750$~K, surface gravity $\log g$\,=\,3.5, and metallicity $[M/H]$\,=\,$-0.25$. The mean parameters of the resulting line mask are $\lambda_0=5067$~\AA, $z_0=1.21$, and $d_0=0.46$. The same parameters were used for normalising the Stokes $I$ and $V$ LSD weights. The LSD profiles were calculated with a 1.2~\kms\ step for the velocity range $\pm$\,60~\kms.

The LSD analysis of the SOFIN observations of \peg\ allowed us to achieve a  $S/N$ gain of 30--40, obtaining a definite detection (false alarm probability $<$\,$10^{-5}$) of the circular polarisation signatures for all but one Stokes $V$ spectrum (phase 0.035 for epoch 2008.7). The resulting mean Stokes $V$ profiles are compatible with the marginal polarisation seen in individual lines (see Fig.~\ref{fig:lsd-sp}), but have a much higher quality, making them suitable for detailed modelling.

\subsection{Mean longitudinal magnetic field}

In Table~\ref{tbl:obs} we report the maximum amplitude of the LSD Stokes $V$ profiles and the mean longitudinal magnetic field \bz\ determined from their first moment. We found typical LSD Stokes $V$ amplitudes of the order of 0.1\% and \bz\ in the range 30--250~G, determined with the precision of $\approx$\,5~G. These results are compatible with the previous longitudinal field measurements by \citet{vogt:1980}, which had an accuracy of 100--160~G and did not yield a convincing detection of magnetic field in \peg.

\begin{figure}[!th]
\fifps{8.5cm}{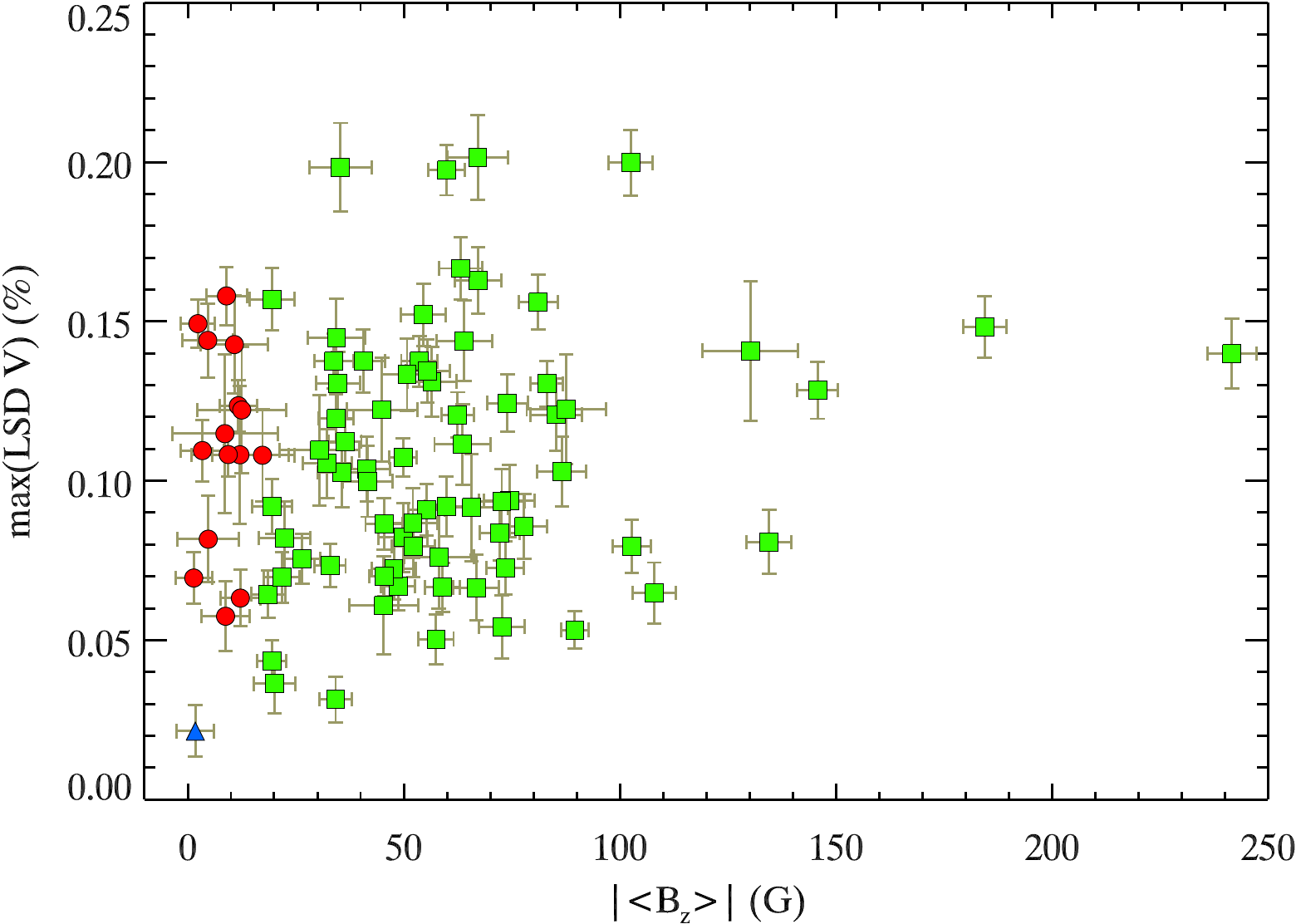}
\caption{Maximum amplitude of the LSD Stokes $V$ profiles as a function of the absolute value of the mean longitudinal magnetic field. The observations for which a polarisation signal is definitely detected and \bz\ is measured at $>$\,3-$\sigma$ significance are shown with squares. Circles correspond to the observations for which \bz\ is not significant, but polarisation signal is still detected. The triangle in the lower-left corner corresponds to a single non-detection of the polarisation signal in the LSD Stokes $V$ profile.}
\label{fig:bz}
\end{figure}

\begin{figure}[!th]
\fifps{8.5cm}{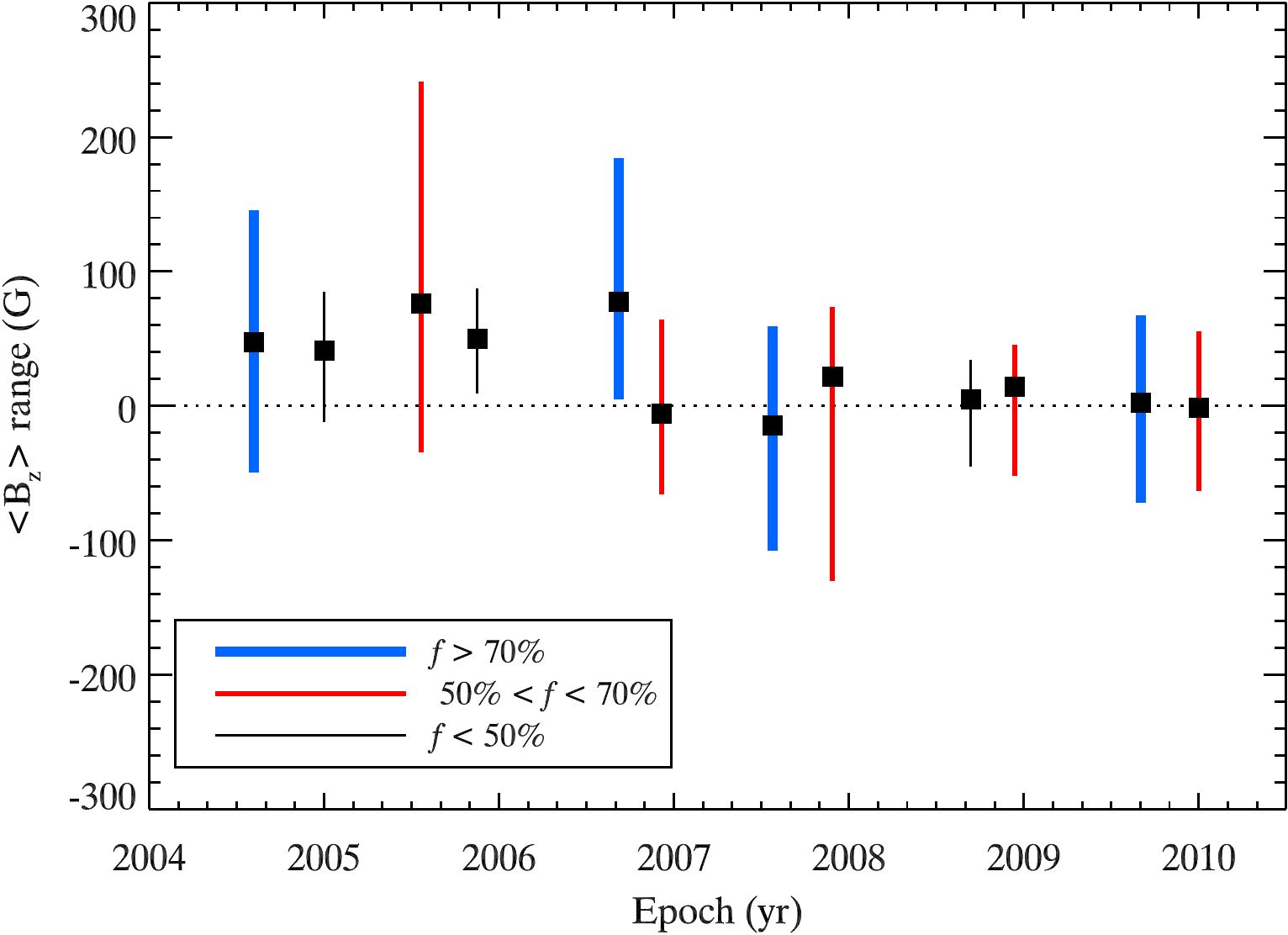}
\caption{Range of longitudinal field variation with rotational phase for different epochs of \peg\ observations. The line thickness is proportional to the phase coverage $f$. The squares show longitudinal field averaged over rotational phase.}
\label{fig:bz-range}
\end{figure}

Despite the ubiquitous definite detections of the circular polarisation signatures, 16 out of 88 LSD profiles of \peg\ do not exhibit a significant mean longitudinal magnetic field. A comparison of \bz\ measurements with the maximum LSD Stokes $V$ amplitude (Fig.~\ref{fig:bz}) shows no obvious correlation. This indicates that, with a few exceptions (e.g. phase 0.531 at epoch 2004.6 illustrated in Fig.~\ref{fig:lsd-sp}), the Stokes $V$ profiles of \peg\ are generally complex and cannot be fully characterised by their first-order moment. Nevertheless, analysis of the longitudinal magnetic field measurements does reveal an intriguing long-term behaviour. As illustrated in Fig.~\ref{fig:bz-range}, the range of the longitudinal field variation with rotational phase systematically decreased from 2004 to 2010. Furthermore, it appears that the phase-averaged value of the longitudinal magnetic field was significantly deviating from 0 during the period from the beginning of our monitoring until the epoch 2006.9. This suggests the presence of large areas of positive magnetic polarity on the visible hemisphere of the star before that epoch and a more symmetric distribution of the field orientation at later epochs.

\section{Self-consistent magnetic and temperature mapping}
\label{zdi}

\subsection{Zeeman Doppler imaging code \inv}
\label{code}

Reconstruction of the distribution of magnetic field and temperature on the surface of \peg\ was carried out with the help of our new ZDI code \inv. This inversion software represents a further development of the \invb\ code \citep{piskunov:2002a,kochukhov:2002c}, previously applied for mapping magnetic geometries and chemical spots on Ap stars \citep{kochukhov:2004d,kochukhov:2010,luftinger:2010}. Both codes incorporate full treatment of the polarised radiative transfer using a modified Diagonal Element Lambda-Operator (DELO) algorithm \citep{rees:1989,piskunov:2002a}. All four Stokes parameters are calculated simultaneously and fully self-consistently, taking into account Zeeman broadening and splitting of spectral lines in the intensity spectra on the one hand, and attenuation of the polarised profiles due to temperature spots on the other hand. Some aspects of this approach have already been incorporated in the ZDI of cool stars in the preliminary studies of \peg\ \citep{carroll:2007,carroll:2009a,kochukhov:2009d}. However, the majority of recent applications of ZDI \citep[e.g.][]{skelly:2010,waite:2011,marsden:2011} still systematically neglect effects of temperature spots in the magnetic inversions. As demonstrated by the numerical tests of \citet{rosen:2012}, the lack of self-consistency in temperature and magnetic mapping can lead to severe artefacts if cool spots coincide with major concentrations of the magnetic flux.

Calculation of the local Stokes parameter profiles by \inv\ is based on a prescribed line list of atomic and molecular lines, normally obtained from the {\sc vald} and {\sc marcs} databases, and a grid of model atmospheres. In the present study of \peg\ we employed a grid consisting of 18 {\sc marcs} model atmospheres covering a $T_{\rm eff}$ range of 3000--5750~K with a 100--250~K step in temperature. The Stokes $IQUV$ profiles and continuum intensities $I_{\rm c}$ are calculated for the temperature of a given stellar surface element by quadratic interpolation between three sets of model spectra corresponding to the nearest points in the model atmosphere grid. This allows an accurate semi-analytical computation of the derivatives with respect to temperature. On the other hand, the derivatives with respect to the three magnetic field vector components are evaluated numerically, using a simple one-sided difference scheme. 

The local profiles are convolved with a Gaussian function to take into account instrumental and radial-tangential macroturbulent broadening, Doppler shifted, and summed for each rotational phase on the wavelength grid of observed spectra. The resulting Stokes parameter profiles are normalised by the phase-dependent continuum flux and compared with observations. In calculating the goodness of the fit we approximately balance the contributions of each Stokes parameter by scaling the respective chi-square terms by the inverse of the mean amplitude of the corresponding Stokes profiles. Test inversions by \citet{kochukhov:2002c} and \citet{rosen:2012} verified simultaneous reconstruction of the magnetic and starspot maps using this weighting scheme. This approach is also routinely used in the reconstruction of Ap-star magnetic fields from four Stokes parameter observations \citep{kochukhov:2004d,kochukhov:2010}. In the study of \peg\ the contributions of the Stokes $I$ and $V$ spectra were weighted as 1:30. Inversion results are not sensitive to the exact choice of the relative weighting of Stokes $I$ and $V$.

The iterative adjustment of the surface maps for matching available observations is accomplished by means of the Levenberg-Marquardt optimisation algorithm, which enables convergence in typically 10--20 iterations starting from a homogeneous temperature distribution and zero magnetic field. The \inv\ code is optimised for execution on massively-parallel computers using MPI libraries. We refer the reader to \citet{kochukhov:2012} for further details on the numerical methods and computational techniques employed in \inv.

As shown numerically by \citet{donati:1997a} and \citet{piskunov:2002a}, and analytically by \citet{piskunov:2005}, ZDI with only Stokes $I$ and $V$ spectra is an intrinsically ill-posed problem, requiring the use of a regularisation or penalty function to reach a stable and unique solution. The reconstruction of temperature in \inv\ is regularised using the Tikhonov method, very similar to many previous DI studies of cool active stars \citep[e.g.][]{piskunov:1993} and our recent temperature mapping of \peg\ with the same SOFIN data \citep{hackman:2012}. In this approach the code seeks a surface distribution with a minimum contrast between adjacent surface elements. 

For the magnetic field reconstruction, \inv\ offers a choice of either directly mapping the three magnetic vector components and applying the Tikhonov regularisation individually to the radial, meridional, and azimuthal magnetic component maps \citep{piskunov:2002a}, or expanding the field into a spherical harmonic series \citep{donati:2006b}. In the latter formulation of the magnetic inversion problem, the free parameters optimised by the code are the spherical harmonic coefficients $\alpha_{\ell m}$, $\beta_{\ell m}$, and $\gamma_{\ell m}$, giving the amplitudes of the poloidal and toroidal terms for a given angular degree $\ell$ and azimuthal number $m$ (see Eqs. 2--8 in \citealt{donati:2006b}). In particular, the radial field is determined entirely by the poloidal contribution (coefficient $\alpha$), whereas both poloidal and toroidal components (coefficients $\beta$ and $\gamma$, respectively) are contributing to the meridional and azimuthal field. In this description of the stellar magnetic field topology through a spherical harmonic expansion, the role of regularisation is played by a penalty function equal to the sum of squares (magnetic energies) of the harmonic coefficients weighted by $\ell^2$. This prevents the code from introducing high-order modes not justified by the observational data. 

Expansion in spherical harmonics has the advantage of automatically satisfying the divergence-free condition for a magnetic field. It also allows us to conveniently characterise the relative contributions of different components to the stellar magnetic field topology and to study the evolution of these contributions with time. Therefore, for the magnetic mapping of \peg\ we adopted the spherical harmonic approach, choosing $\ell_{\rm max}=10$ which corresponds to a total of 360 independent magnetic field parameters.

Similar to other quantities obtained by solving a regularised ill-posed problem, the spherical harmonic coefficients cannot be attributed formal error bars because, in addition to the observational data, the individual modes are constrained by the penalty function. Therefore, their formal error bars are strongly dependent on the choice of regularisation parameter, especially for the high-$\ell$ modes. This is why here, as well as in all previous ZDI studies employing spherical harmonic expansion \citep[e.g.][]{donati:2006b,skelly:2010,fares:2012}, no formal uncertainty estimates are derived for the harmonic coefficients.

As inferred from the previous numerical tests of ZDI inversions \citep{donati:1997a,kochukhov:2002c,rosen:2012}, reconstruction of the magnetic field distributions from the Stokes $I$ and $V$ data alone suffers from some ambiguities. The most important problem is a cross-talk between the radial and meridional field components at low latitudes (i.e. for the regions below the stellar equator for the rotational geometry of \peg). On the other hand, the harmonic regularisation or parameterisation of the magnetic field maps partly alleviates this problem \citep{donati:2001a,kochukhov:2002c}, because in this case the inversion code is not allowed to vary the three magnetic vector components arbitrarily. Instead, the meridional and azimuthal components are coupled via the $\beta$ and $\gamma$ harmonic coefficients which avoids the cross-talk between the radial and meridional fields, at least for the low-$\ell$ values.

\subsection{Reconstructing temperature spots from LSD profiles}
\label{lsdt}

\begin{figure}[!th]
\firrps{7.5cm}{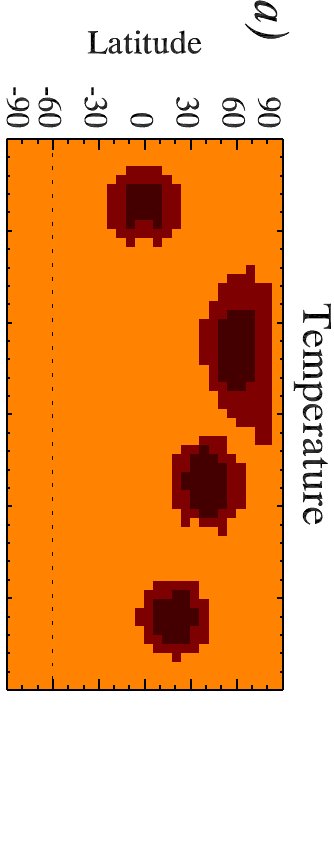}
\firrps{7.5cm}{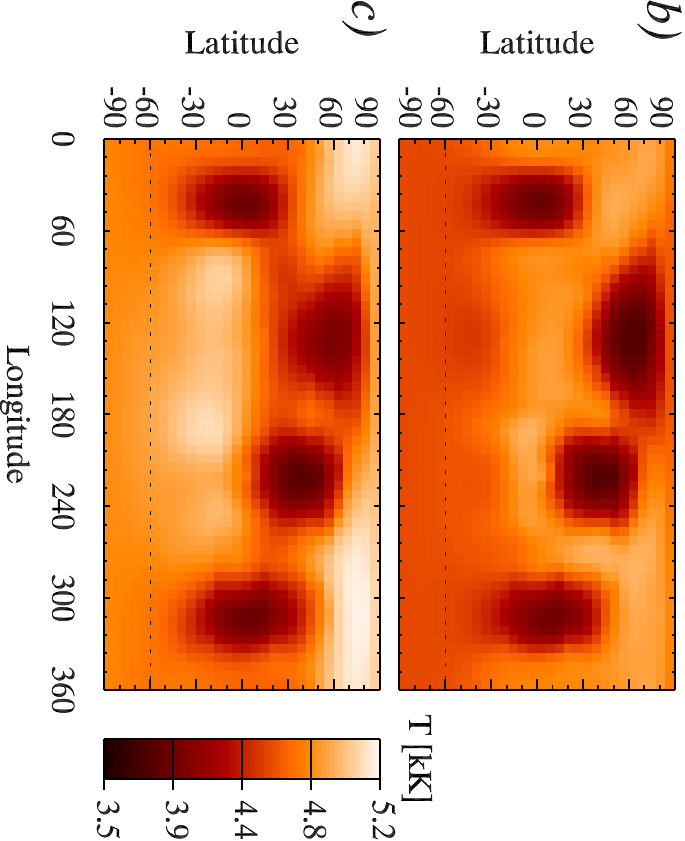}
\caption{Comparison of temperature mapping from a real line and from LSD profiles. {\bf a)} True temperature distribution. {\bf b)} DI map reconstructed from the simulated observations of a real spectral line. {\bf c)} Temperature reconstruction from the simulated LSD profiles assuming mean-line parameters.}
\label{fig:di-test}
\end{figure}

Most previous ZDI modelling of the Stokes $I$ and $V$ LSD profiles of late-type active stars relied on simple analytical formulas, often a Gaussian approximation, to represent the local intensity and polarisation profiles. These coarse line-profile models have either entirely neglected dependence of the local line intensity on temperature \citep[e.g.][]{petit:2004a} or used \textit{disk-integrated} spectra of suitably chosen cool and hot slowly rotating template stars to represent \textit{local} contribution of the spot and photosphere, respectively \citep[e.g.][]{donati:2003}. 

Here we develop a more sophisticated approach to interpret the LSD profile time series. We aim to describe the LSD Stokes $I$ and $V$ spectra with self-consistent polarised radiative transfer calculations, based on realistic model atmospheres, for a fictitious spectral line with average atomic parameters. Comprehensive investigation of the properties of LSD profiles by \citet{kochukhov:2010a} showed that this approach yields an accurate description of the mean circular polarisation line shapes for the magnetic field strengths below $\approx$\,2~kG. For stronger fields, details of the LSD Stokes $V$ profile shape cannot be reproduced, although the LSD profile moments still yield correct longitudinal magnetic field. According to previous studies, magnetic fields found on the surfaces of RS~CVn- and BY~Dra-type active stars are safely within this limit. On the other hand, analyses of Ap/Bp stars hosting much stronger fields \citep[e.g.][]{wade:2000b,silvester:2012} proves that the LSD line-averaging method itself does not systematically underestimate magnetic field strength even beyond the weak-field regime.

At the same time, \citet{kochukhov:2010a} found that a response of the LSD Stokes $I$ profile to the variation of chemical composition differs from that of any real spectral line, which prevents an accurate mapping of the chemical spots in Ap stars using LSD profiles \citep{folsom:2008}. One might suspect that reconstruction of the temperature inhomogeneities from the LSD spectra is similarly limited. It is clear that the temperature sensitivity of an average profile, composed of an essentially random mixture of thousands of lines with diverse temperature responses, cannot be known as well as the one for a set of judiciously chosen individual spectral lines with accurate atomic parameters. On the other hand, typical Stokes $I$ atomic line profile variations in active stars are dominated by the continuum brightness effect, which influences all lines in a similar way \citep{vogt:1983a,unruh:1995}. Adding to this a greatly enhanced $S/N$ of the LSD spectra, one might hope to achieve a reasonable reconstruction of temperature spots treating the LSD intensity profiles as a real spectral line. To clarify whether this is indeed possible, we have carried out numerical tests of temperature inversions based on simulated LSD spectra. 

First, we established mean-line parameters appropriate for the LSD line mask applied to the SOFIN observations of \peg. Among the 1580 spectral lines included in this mask, \ion{Fe}{i} is the most common ion (about 40\% of all lines). Thus, we adopted the \ion{Fe}{i} identification for a fictitious line representing LSD profiles and obtained the preliminary set of atomic parameters necessary for spectrum synthesis by combining the average excitation energy of the lower level, oscillator strength, and broadening constants for this ion with the mean wavelength and Land\'e factor (5067~\AA\ and 1.21, respectively) for the entire line list. Then we adjusted the oscillator strength and van der Waals damping constant by fitting the synthetic LSD profiles derived from a theoretical spectrum covering the same wavelength interval as real observations. These calculations adopted the stellar atmospheric parameters discussed below (Sect.~\ref{atm}) and were performed for the \vs\ of \peg\ and for the non-rotating star with the same parameters. In both cases we found that single-line calculations provide an excellent fit to the LSD profiles obtained from the synthetic spectra.

At the next step, we simulated spectra of a star with temperature inhomogeneities, using a photospheric temperature of 4750~K and a spot temperature of 3700~K. The Stokes $I$ spectra for the entire SOFIN wavelength range were computed for 10 equidistant rotational phases, assuming $i$\,=\,60\degr, \vs\,=\,23~\kms, and no magnetic field. A surface temperature map consisting of four circular spots located at different latitudes (see Fig.~\ref{fig:di-test}a) was employed in these calculations. These theoretical spectra were processed with our LSD code in exactly the same way as real stellar observations. The resulting set of synthetic LSD profiles was interpreted with \inv\ using the line parameters established above. For comparison, we also reconstructed a temperature map from the direct forward calculations for the same spectral line.

A comparison of the DI maps inferred from the simulated data with the true temperature distribution is presented in Fig.~\ref{fig:di-test}. The reconstructions from the simulated observations of the real spectral line and from the simulated LSD profiles are generally successful in recovering the properties of cool spots. However, the temperature DI using LSD profiles shows a tendency to produce spurious hot surface features in which the temperature is overestimated by 200--300~K. We emphasise that this problem appears entirely due to a loss of information about real temperature sensitivity of the local profiles of individual spectral lines. Fortunately, this bias does not significantly interfere with the reconstruction of large low-temperature spots, although their contrast seems to be systematically underestimated. Thus, we conclude that the temperature DI based on a single-line radiative transfer approximation of the LSD profiles is a viable approach to mapping the distribution of cool spots on the surfaces of active stars. At the same time, the reliability of hot features recovered with this method is questionable.

Limitations of LSD-based temperature inversions partly explain why some of the temperature maps presented here differ from those in our previous analysis \citep{hackman:2012}. The differences are mainly caused by the use here of LSD profiles, while our previous analysis was based on individual lines. While individual lines will give a more accurate result, this approach suffers from a lower S/N. At the same time, neglect of the Zeeman broadening and intensification in earlier temperature inversions may also contribute to the discrepancy with the present results. In general, both of these effects will influence the spot temperatures and latitudes.

\subsection{Stellar parameters}
\label{atm}

Different sets of atmospheric parameters have been deduced in previous studies of \peg. The most thorough model atmosphere analyses were those by \citet{berdyugina:1998} and \citet{ottmann:1998}. The first paper suggested \teff\,=\,4600~K, $\log g$\,=\,3.2, and $[M/H]$\,=\,$-0.4$, whereas the second paper has inferred \teff\,=\,4800~K, $\log g$\,=\,3.65, and $[Fe/H]$\,=\,$-0.24$. A comparison of the average SOFIN observations with the theoretical spectra computed using the {\sc synth3} code \citep{kochukhov:2007d} and {\sc marcs} model atmospheres supports the latter set of stellar parameters. Therefore, we adopted a {\sc marcs} model atmosphere with \teff\,=\,4750~K, $\log g$\,=\,3.5, and $[M/H]$\,=\,$-0.25$ to represent an unspotted star. This temperature was also used as a starting guess for DI. All spectrum synthesis calculations assumed a microturbulent velocity of $\xi_{\rm t}$\,=\,2~\kms\ and a radial-tangential macroturbulent broadening of $\zeta_{\rm t}$\,=\,4~\kms. These parameters are compatible with the values determined in previous studies. We note that the synthetic \ion{Fe}{i} line approximating the LSD profiles of \peg\ is weak ($W_\lambda$\,=\,40~m\AA) and, therefore, relatively insensitive to the choice of $\xi_{\rm t}$.

Based on previous studies \citep{berdyugina:1998,frasca:2008}, we adopted an inclination of the stellar rotational axis $i=60\degr$. Using the three data sets with the best phase-coverage (epochs 2004.6, 2007.6, and 2009.7), we found that the best fit to the Stokes $I$ LSD profiles is achieved for \vs\,=\,$23\pm0.5$~\kms. This value of the projected rotational velocity is consistent with \vs\,=\,$22.6\pm0.5$~\kms\ determined by \citet{berdyugina:1998} and used in the recent DI studies by \citet{lindborg:2011} and \citet{hackman:2012}.

A latitude-dependent differential rotation can be incorporated in the ZDI with \inv. This effect was ignored by previous spectroscopic DI investigations of \peg\ and was theoretically predicted to be negligible in the rapidly rotating stars with deep convective envelopes \citep{kitchatinov:1999}. On the other hand, photometric analyses occasionally claimed detection of the differential rotation in \peg\ \citep{henry:1995,siwak:2010,roettenbacher:2011}, yet giving contradictory results, mainly because of the difficulty of constraining the spot latitudes in light curve modelling. However, even the largest value of the differential rotation coefficient reported in the literature $k=0.0245$ \citep{siwak:2010} is substantially smaller than the solar value of $k=0.19$ and corresponds to an insignificant surface shearing for the time span of most of our SOFIN data sets. Therefore, ZDI analysis of \peg\ was carried out assuming no differential rotation.

\section{Magnetic field topology of \peg}
\label{res}

\begin{figure*}[!t]
\centering
\twofig{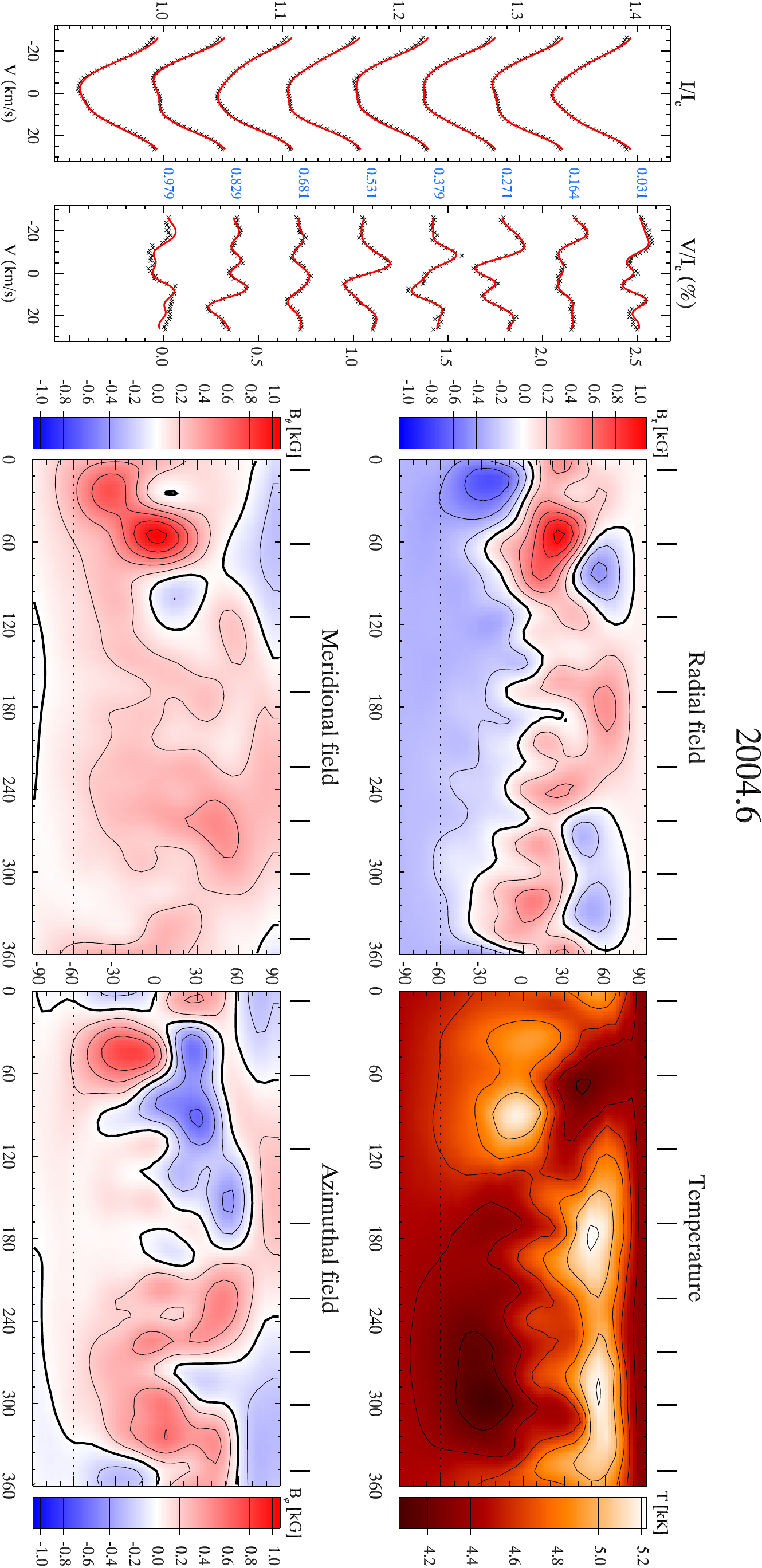}{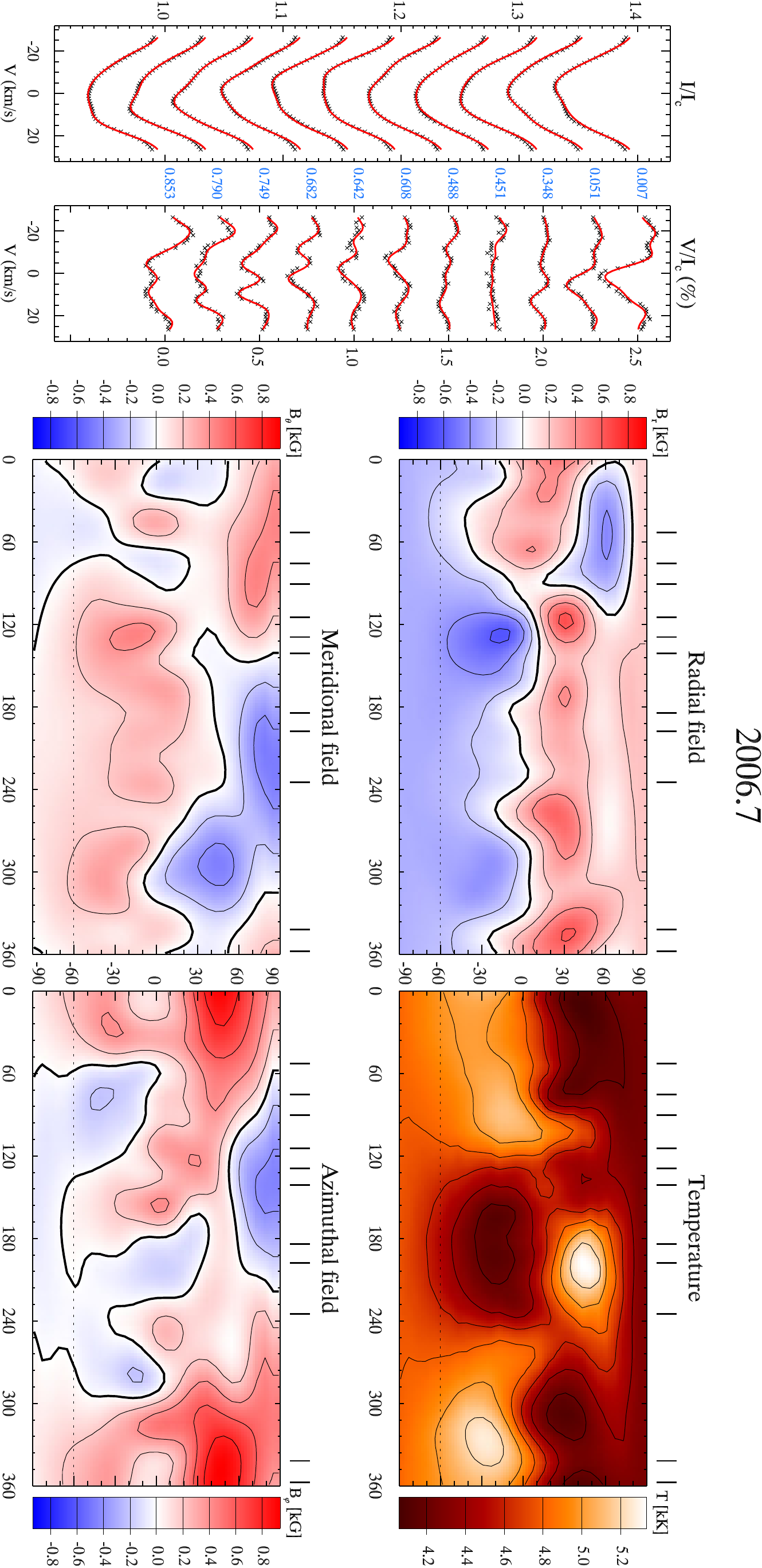}
\caption{Magnetic field topology and temperature distribution of \peg\ at epochs 2004.6 and 2006.7. For each epoch the two columns on the left  compare the observed Stokes $I$ and $V$ LSD profiles (\textit{symbols}) and the theoretical spectrum synthesis fit (\textit{solid lines}). Spectra corresponding to different rotational phases are shifted vertically. The scale is given in \% for Stokes $V$ and in units of the continuum intensity for Stokes $I$. The four rectangular maps represent surface distributions of the radial, meridional, and azimuthal magnetic field components and corresponding temperature reconstructed for each epoch. 
The contours are plotted with a step of 0.2~kG over the magnetic maps and with a step of 200~K over the temperature distributions. 
The thick contour lines indicate where the field changes sign in the magnetic maps.
The vertical bars above each rectangular panel indicate rotational phases of individual observations. The rotational phase runs from right to left on all maps.
}
\label{fig:map_a}
\end{figure*}

\begin{figure*}[!t]
\centering
\twofig{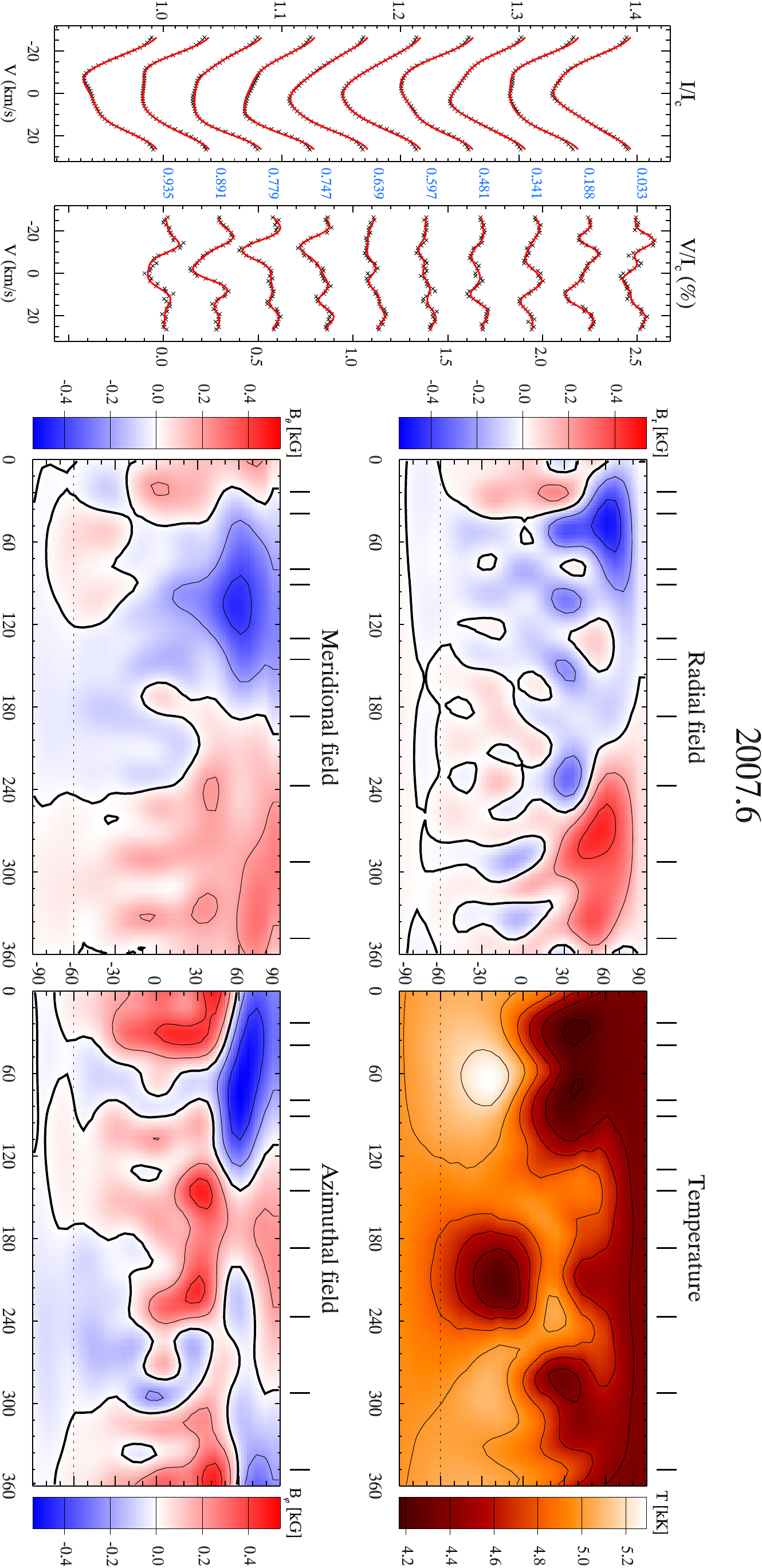}{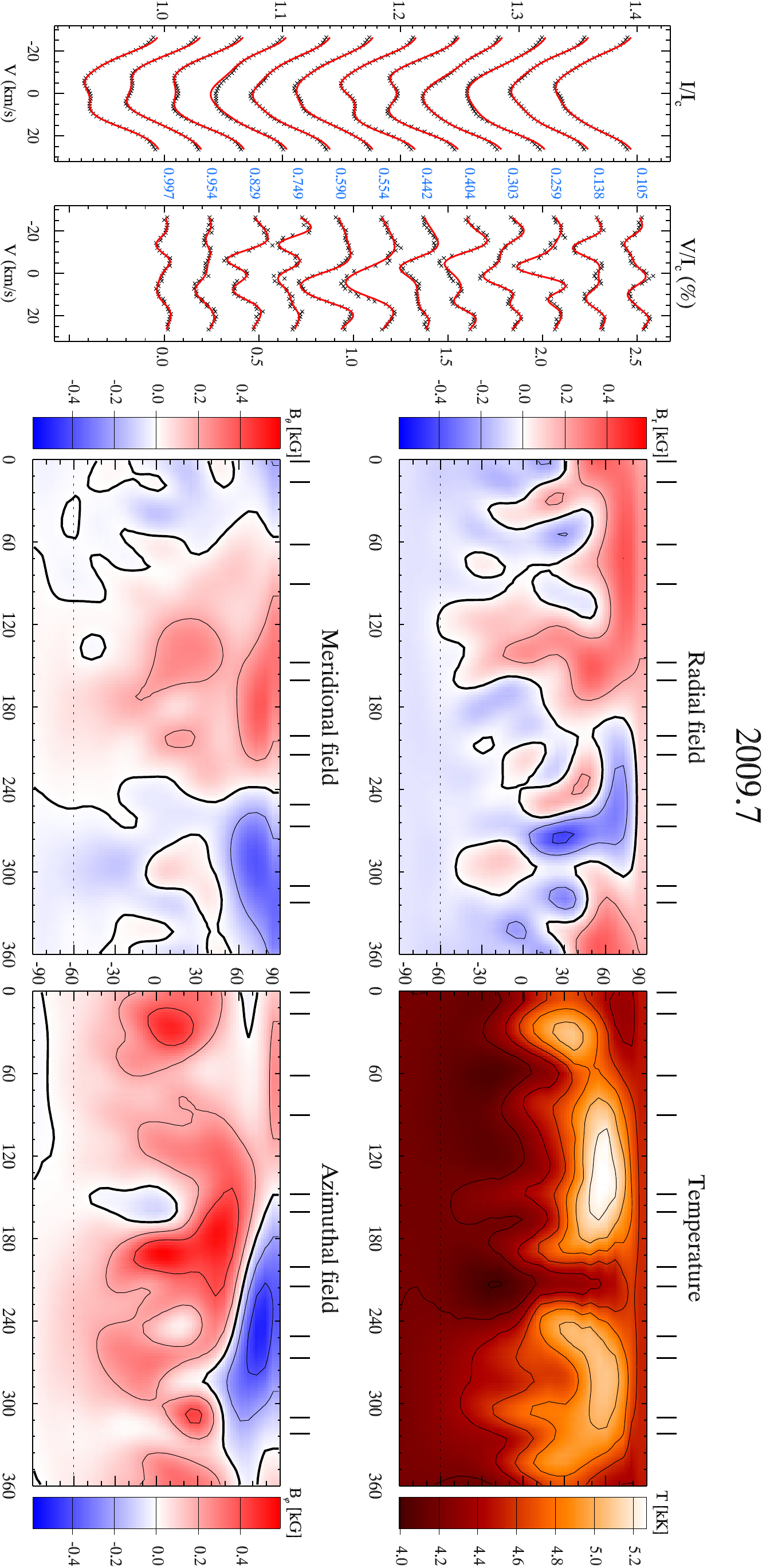}
\caption{Same as Fig.~\ref{fig:map_a} for epochs 2007.6 and 2009.7.
}
\label{fig:map_b}
\end{figure*}

Results of the simultaneous reconstruction of magnetic field topology and temperature distribution for \peg\ are illustrated for the four data sets with best phase-coverage in Figs.~\ref{fig:map_a} and \ref{fig:map_b} and for the remaining epochs in the online Figs.~\ref{fig:map_c}--\ref{fig:map_e}. For each data set we present four rectangular maps, showing the distribution of temperature as well as the radial, meridional, and azimuthal field components. A comparison between observed and computed LSD Stokes $I$ and $V$ profiles is illustrated next to the corresponding surface images. The model profiles generally achieve an adequate fit to observations, with median mean deviations of 0.15\% and 0.01\% for the Stokes $I$ and $V$ LSD profiles, respectively.

\onlfig{
\begin{figure*}[!t]
\centering
\twofig{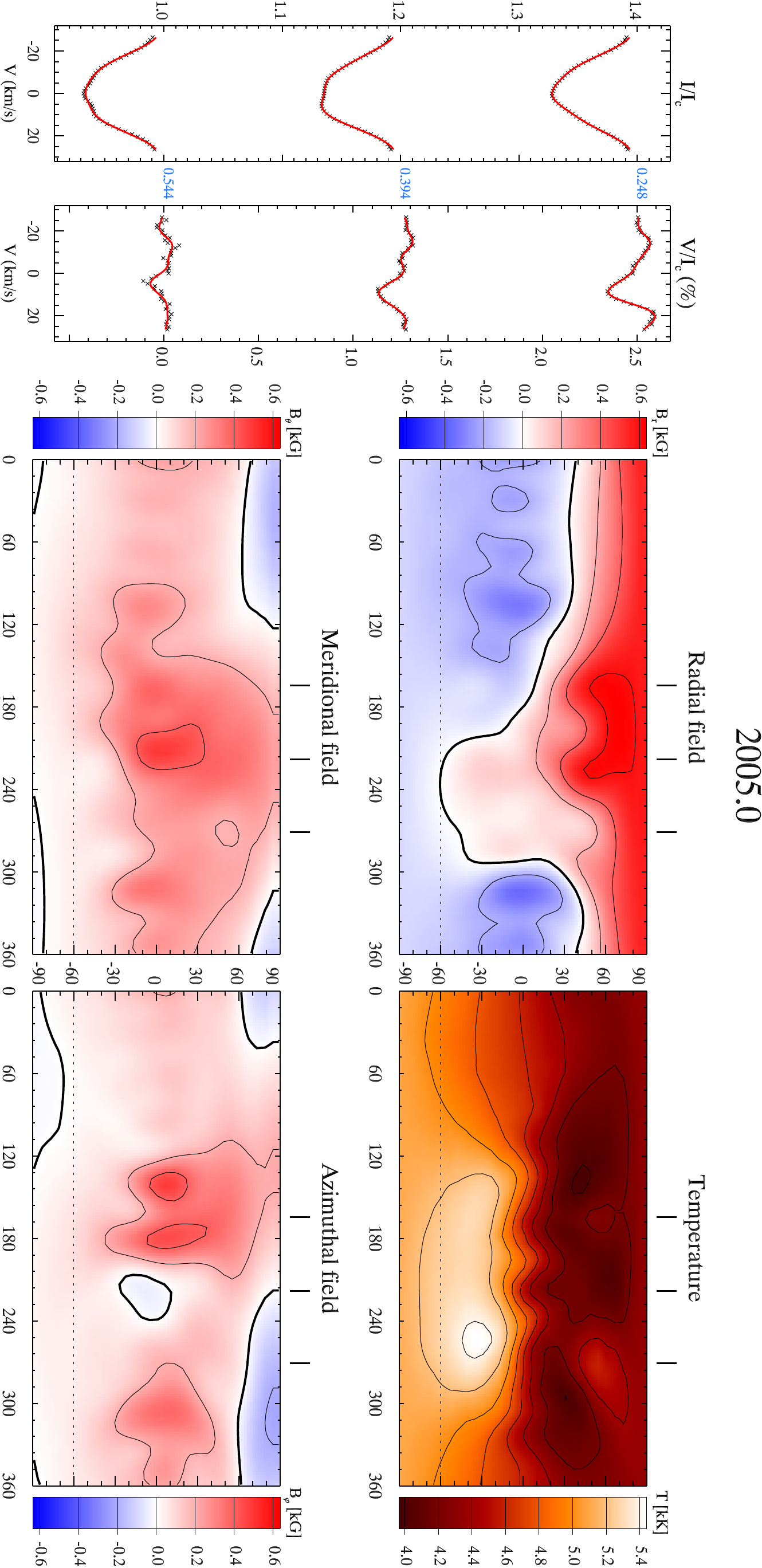}{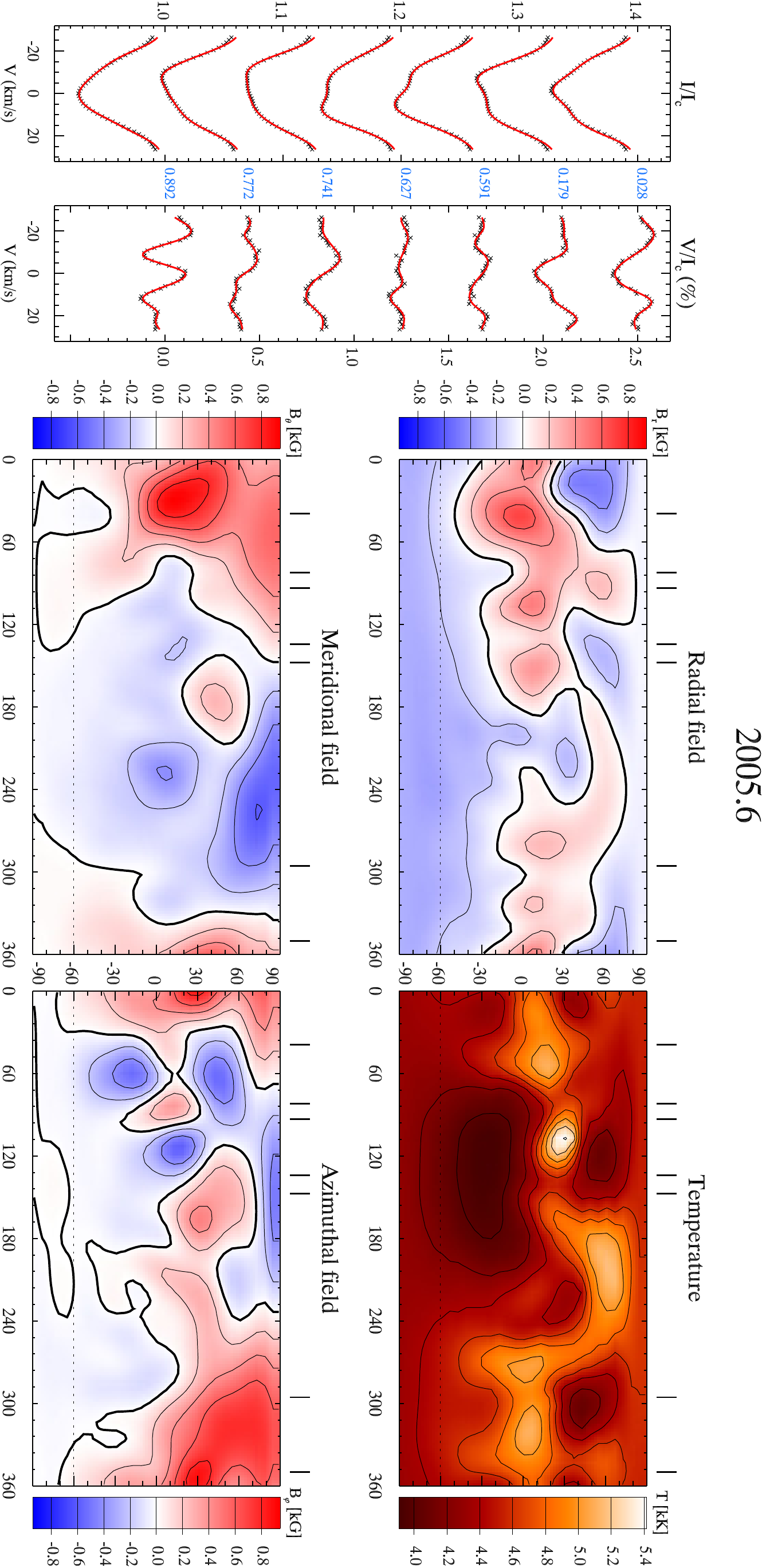}
\caption{Same as Fig.~\ref{fig:map_a} for epochs 2005.0 and 2005.6.
}
\label{fig:map_c}
\end{figure*}
}

\onlfig{
\begin{figure*}[!t]
\centering
\twofig{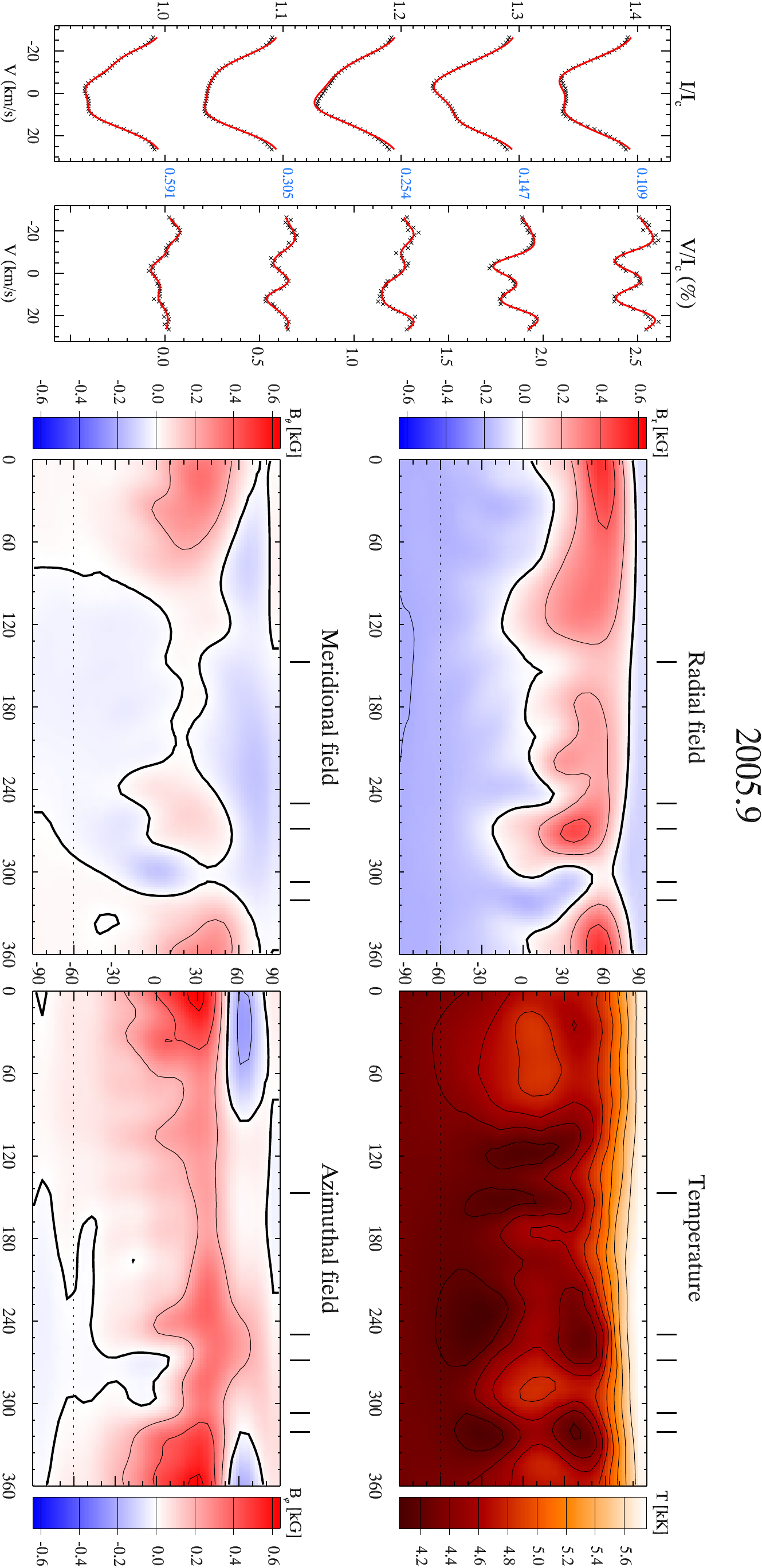}{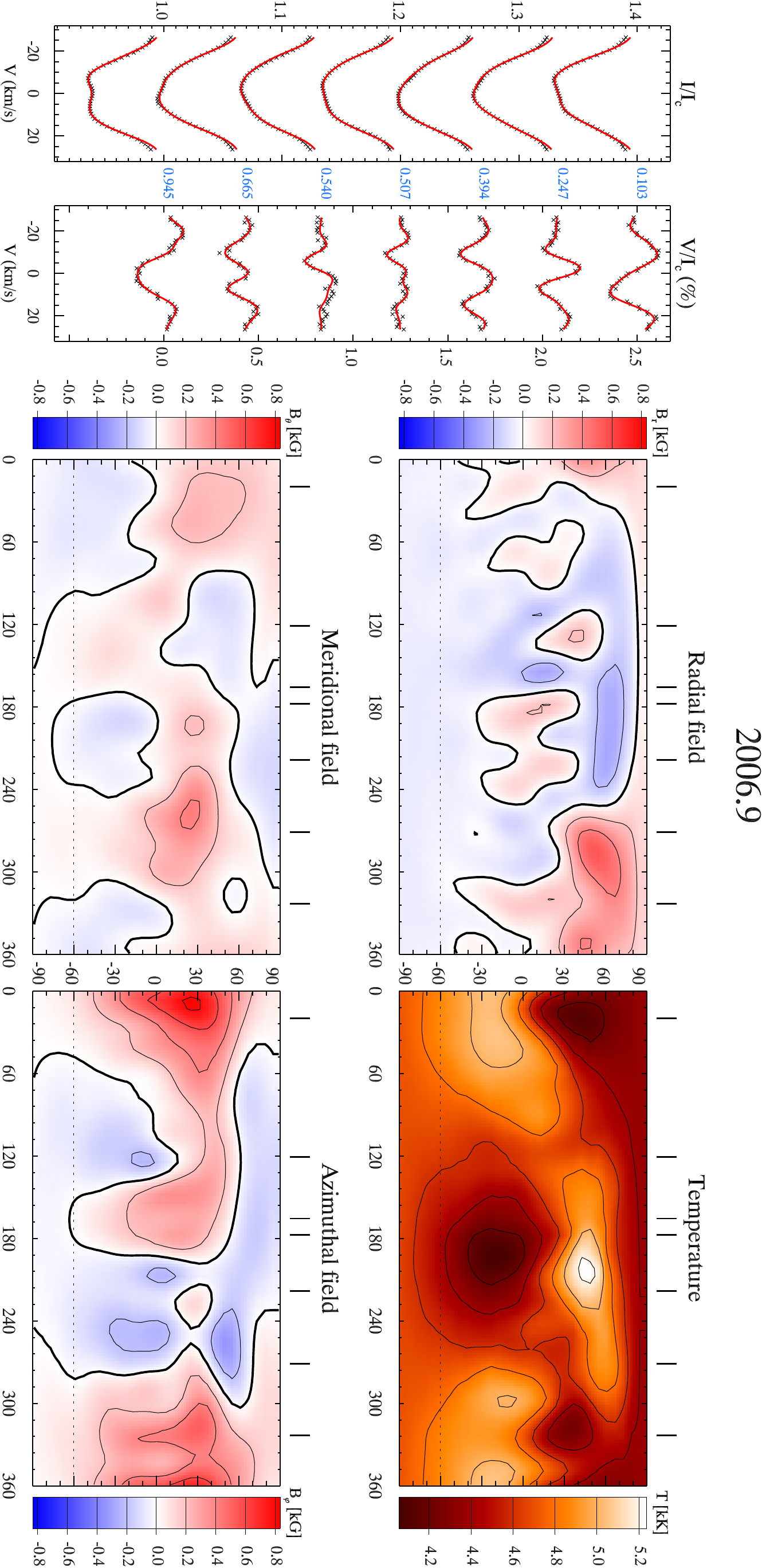}
\caption{Same as Fig.~\ref{fig:map_a} for epochs 2005.9 and 2006.9.
}
\label{fig:map_d}
\end{figure*}
}

\onlfig{
\begin{figure*}[!t]
\centering
\twofig{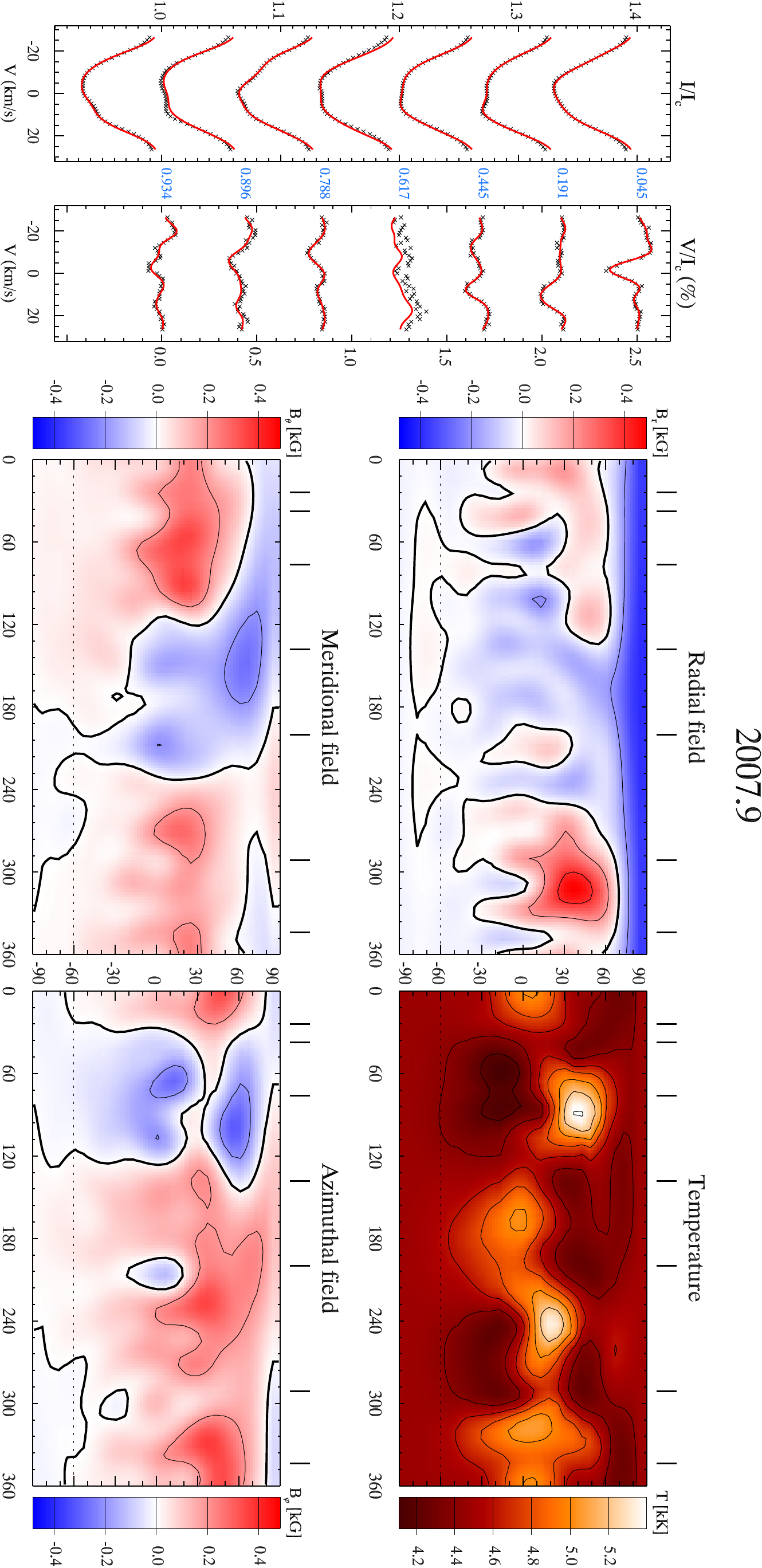}{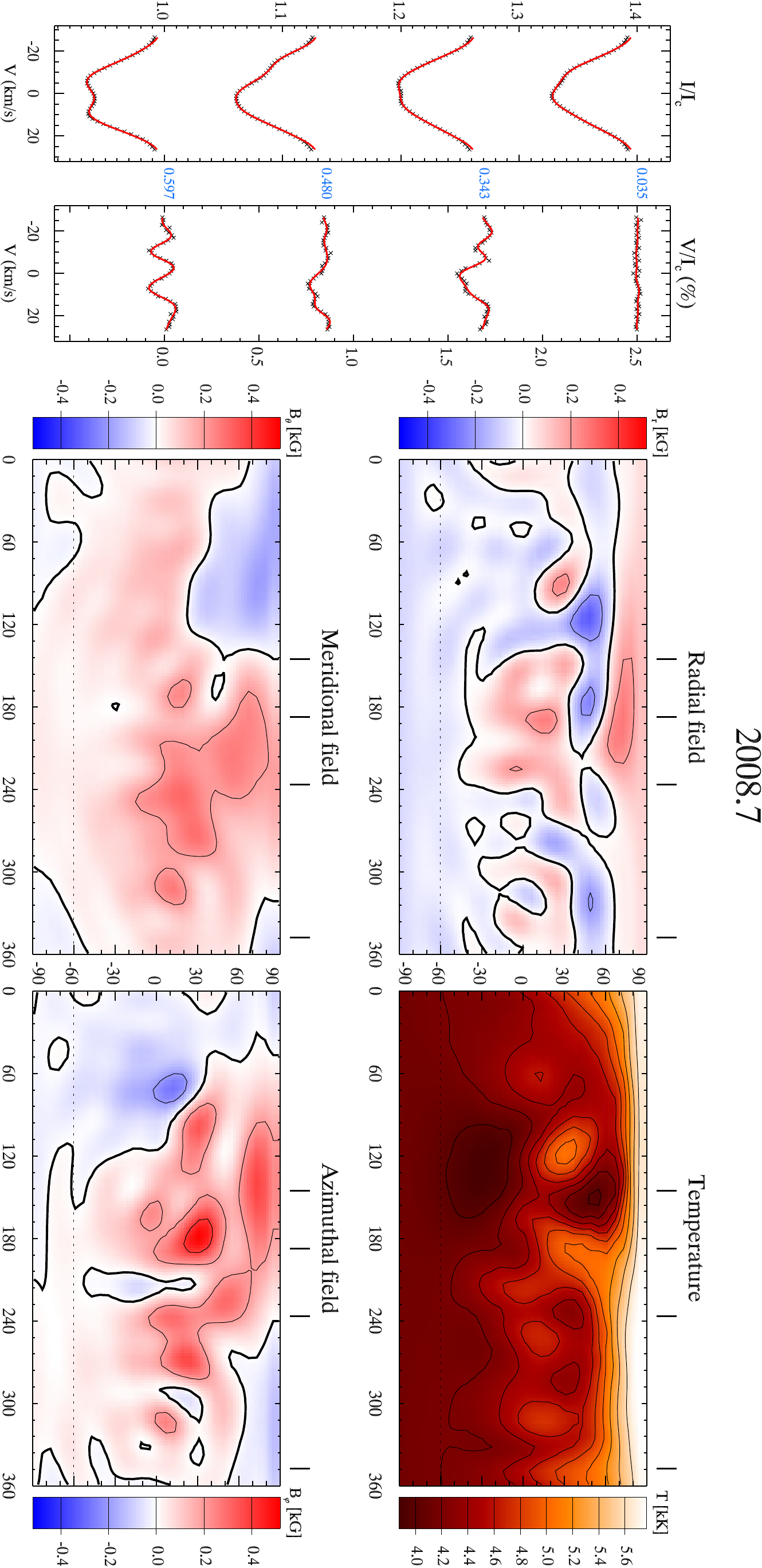}
\caption{Same as Fig.~\ref{fig:map_a} for epochs 2007.9 and 2008.7.
}
\label{fig:map_e}
\end{figure*}
}

\onlfig{
\begin{figure*}[!t]
\centering
\twofig{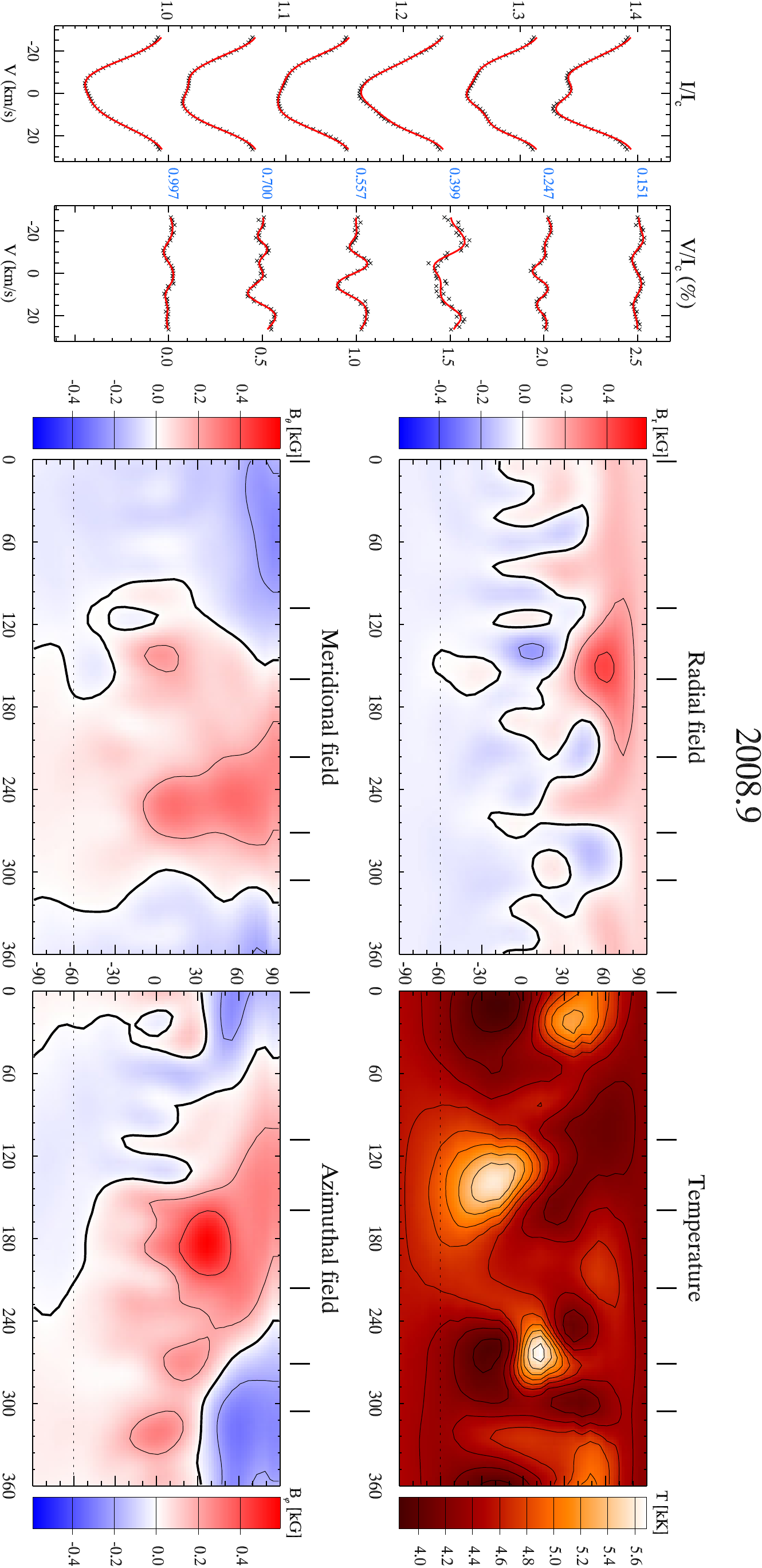}{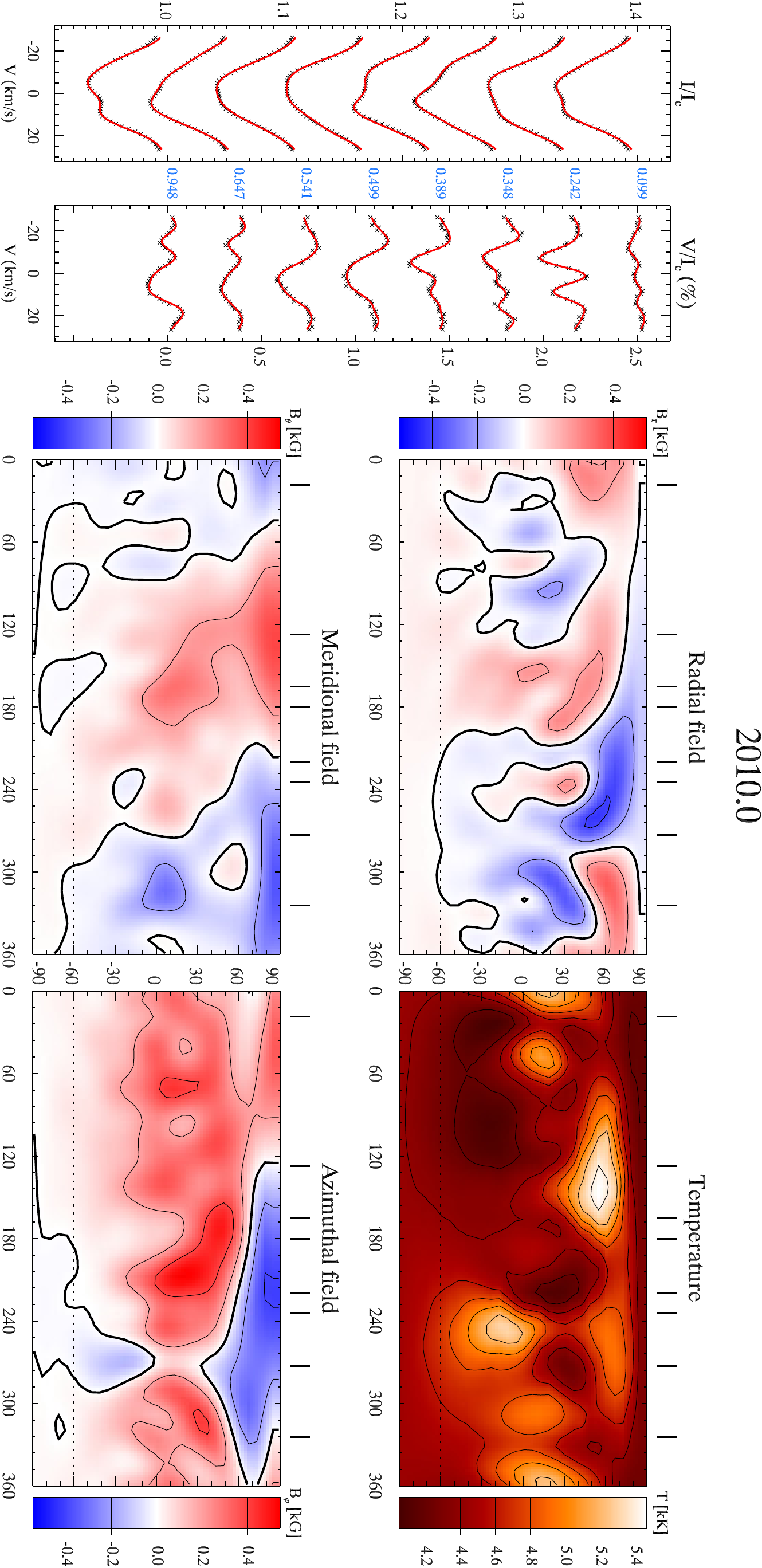}
\caption{Same as Fig.~\ref{fig:map_a} for epochs 2008.9 and 2010.0.
}
\label{fig:map_f}
\end{figure*}
}

The maximum local magnetic field intensity inferred by our ZDI code reaches 1~kG for 2004.6 data set, but is typically on the order of 400--500~G for the later magnetic images. Judging from these results, the field strength on the surface of \peg\ was steadily decreasing from our first observations to later epochs. The field topology appears to be rather complex in 2004--2005, with many localised magnetic spots in all three magnetic vector component maps. As the field became weaker, this configuration evolved into a simpler topology, often dominated by large regions of common field polarity (e.g. epoch 2007.6).

The temperature maps of \peg\ reconstructed from LSD profiles show cool spots with an 800--1000~K contrast relative to the photospheric $T=4750$~K. Occasionally, the inversion code reconstructed spots which are a few hundred K hotter than the photosphere. However, the numerical experiments discussed in Sect.~\ref{lsdt} suggest that these features are not reliable.

Inversions were carried out for all epochs, including the three data sets (2005.0, 2005.9, and 2008.7) with a poor phase-coverage. Temperature maps for these three data sets are dominated by a strong axisymmetric component, which should be regarded as spurious.

\subsection{Average field parameters}

\begin{figure}[!th]
\fifps{8cm}{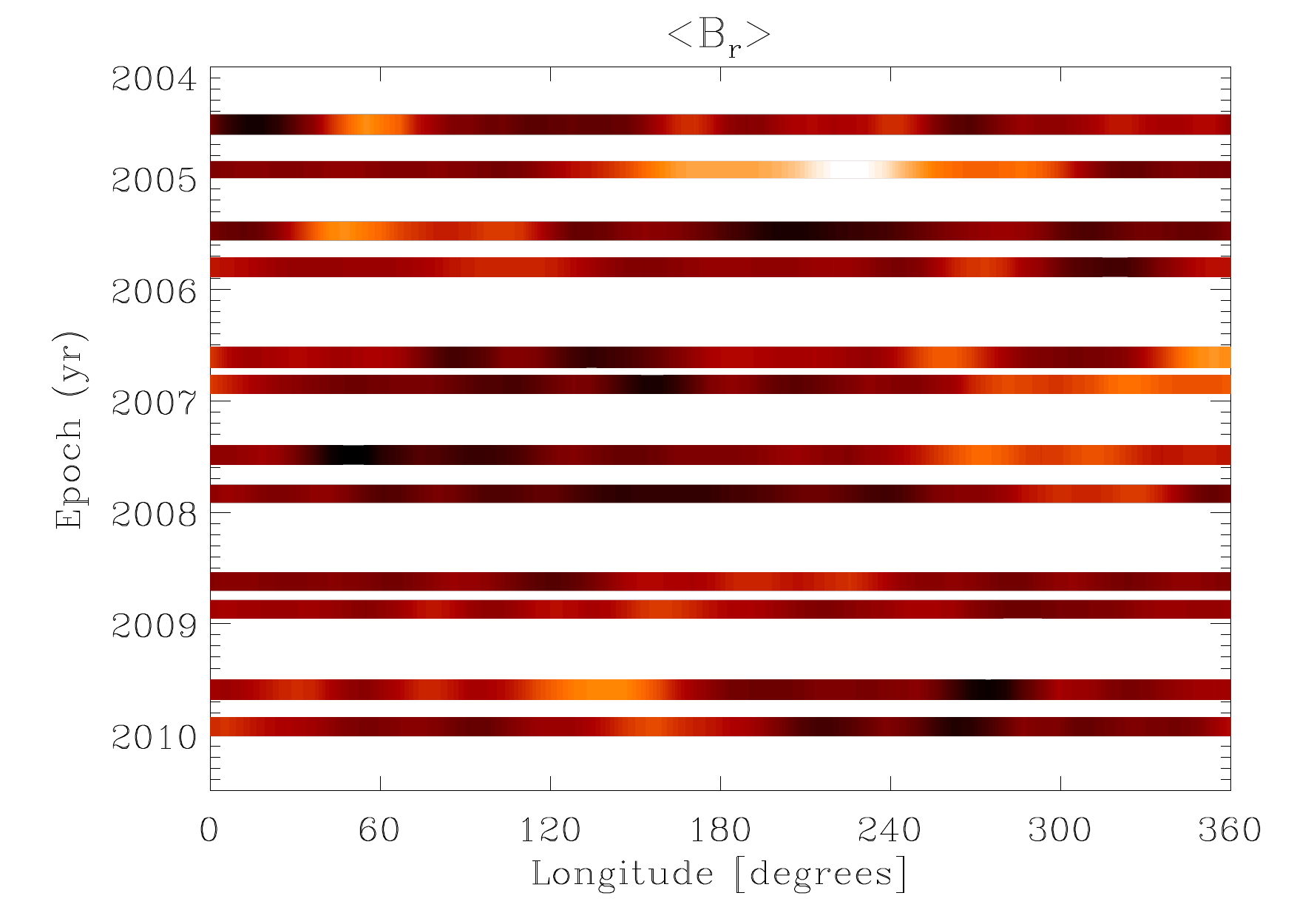}
\fifps{8cm}{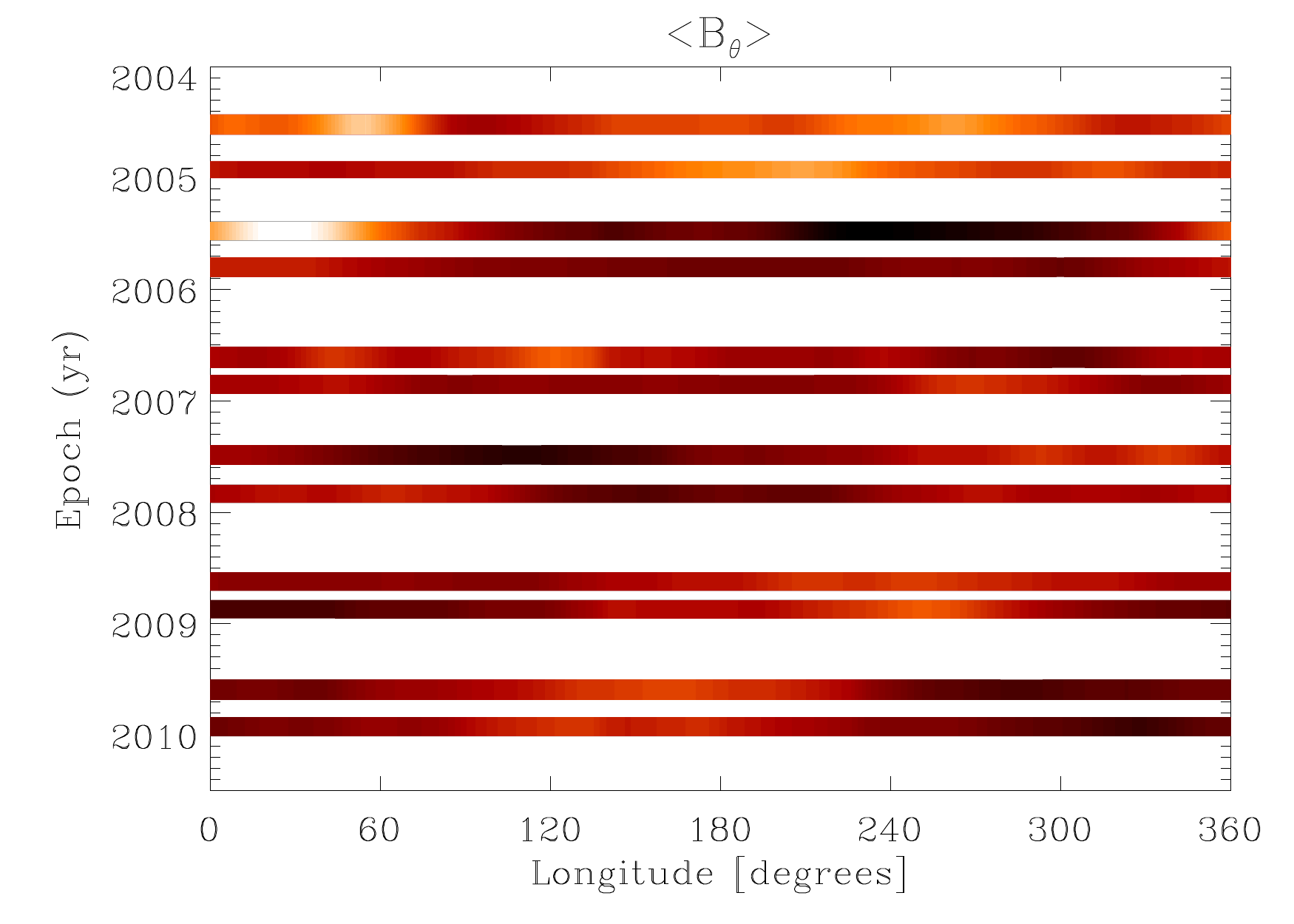}
\fifps{8cm}{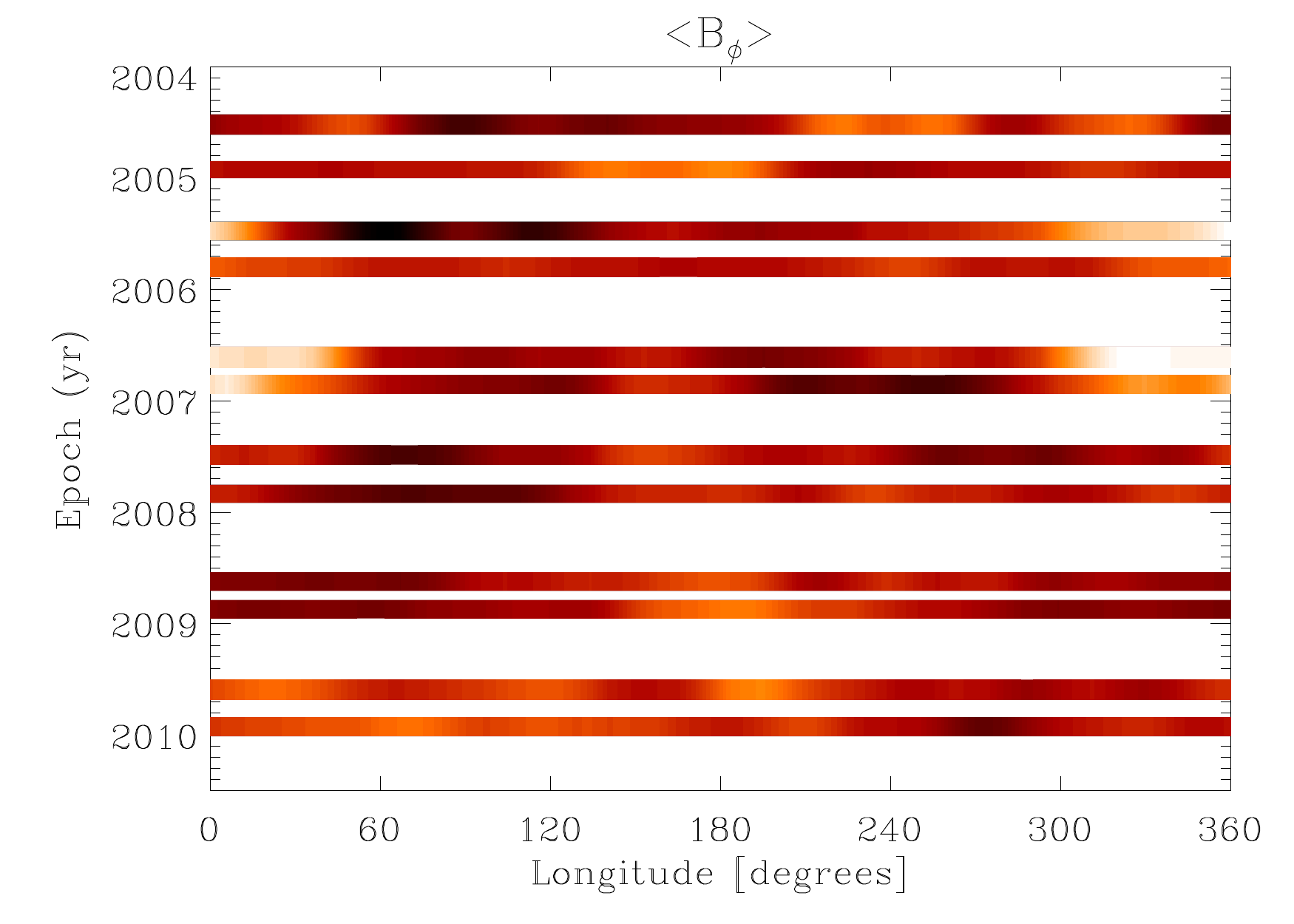}
\caption{Magnetic field components (top: radial field, middle: meridional field, bottom: azimuthal field) averaged over the latitude interval
$-60\degr$ to $90\degr$ from each ZDI map. Each stripe represents the average field from one ZDI map, extended over the time axis to help
the visualisation. The stripes are organised as function of time, according to their observational epoch, running from top to bottom. Dark colours represent negative polarities, bright colours positive ones; in all figures a linear colour table with minimum of $-300$~G and maximum of 300~G has been used.}
\label{fig:phasetime}
\end{figure}

The two independent ZDI reconstructions obtained for the epochs 2006.7 and 2006.9 allow us to investigate short-term changes of the surface distributions. The temperature maps inferred from the observations separated by three months are very similar. Both images are dominated by a large low-latitude cool spot at 180\degr\ longitude and an extended high-latitude spot group in the longitude interval 300\degr\ to 60\degr. The azimuthal field maps, representing the dominant magnetic component for both epochs, are also fairly similar. At the same time, distributions of the weaker radial and meridional field components exhibit significant differences, suggesting field evolution.

Next, to detect possible systematic changes either in the phases of
the surface magnetic field distribution or in the total magnetic field
strength, we calculated different types of averages of the ZDI
maps. First, we averaged over the visible latitude range in each map, to
obtain magnetic field profiles that depend on longitude (or phase)
only. The profiles shown in Fig.~\ref{fig:phasetime} are averaged over the
latitude interval from $-60\degr$ to $90\degr$ for the different magnetic field
components as function of time. This procedure can reveal azimuthal
dynamo waves, that were visible in the surface temperature maps of
this object during 1994--2002 \citep[see][]{lindborg:2011}. There is no evidence
for this type of dynamo wave from the ZDI maps. This agrees
with the results of \citet{hackman:2012}, who reported the
disappearance of the clear drift pattern in the surface temperature
maps during these years. The radial field plot, on the other hand,
reveals a rather abrupt appearance and disappearance of spots of opposite polarities at a certain
phase which are, however, irregular with time. 

Finally, we calculated the root-mean-square values of the magnetic
field over each ZDI map, characterising the overall magnetic field
strength (shown in Fig.~\ref{fig:brms}). Interestingly, all the magnetic
field components are of comparable strength, immediately hinting, from basic dynamo theory,
towards an $\alpha^2$ dynamo operational in the object -- the presence of differential rotation would lead to
the efficient shearing of the poloidal field into a toroidal field, in
which case the azimuthal component would be observed to dominate over
the other components. The radial and meridional components are clearly
decreasing monotonically with time, reaching a minimum
at around the year 2009. After that, the radial field possibly starts rising again. The azimuthal field shows a somewhat
different trend in time: at first it is somewhat weaker than the other
components, slightly increasing during 2004--2007, after which it also
shows a decreasing trend. Similar to the radial component, it seems to start
rising again after 2010. Therefore, judging from the magnetic field strength, it seems plausible that the magnetic activity level of the
star has been declining during the epoch 2004--2009, while the signs of
rising activity can be seen at least in the rms field strength for later epochs.

\subsection{Evolution of harmonic field components}

Magnetic field maps reconstructed by our ZDI code are parameterised in terms of the spherical harmonic coefficients corresponding to the poloidal and toroidal field components. In addition to the analysis of 2D maps, this representation provides another convenient possibility to characterise the field topology and its long-term evolution.

\begin{figure}[!th]
\centering
\fifps{8cm}{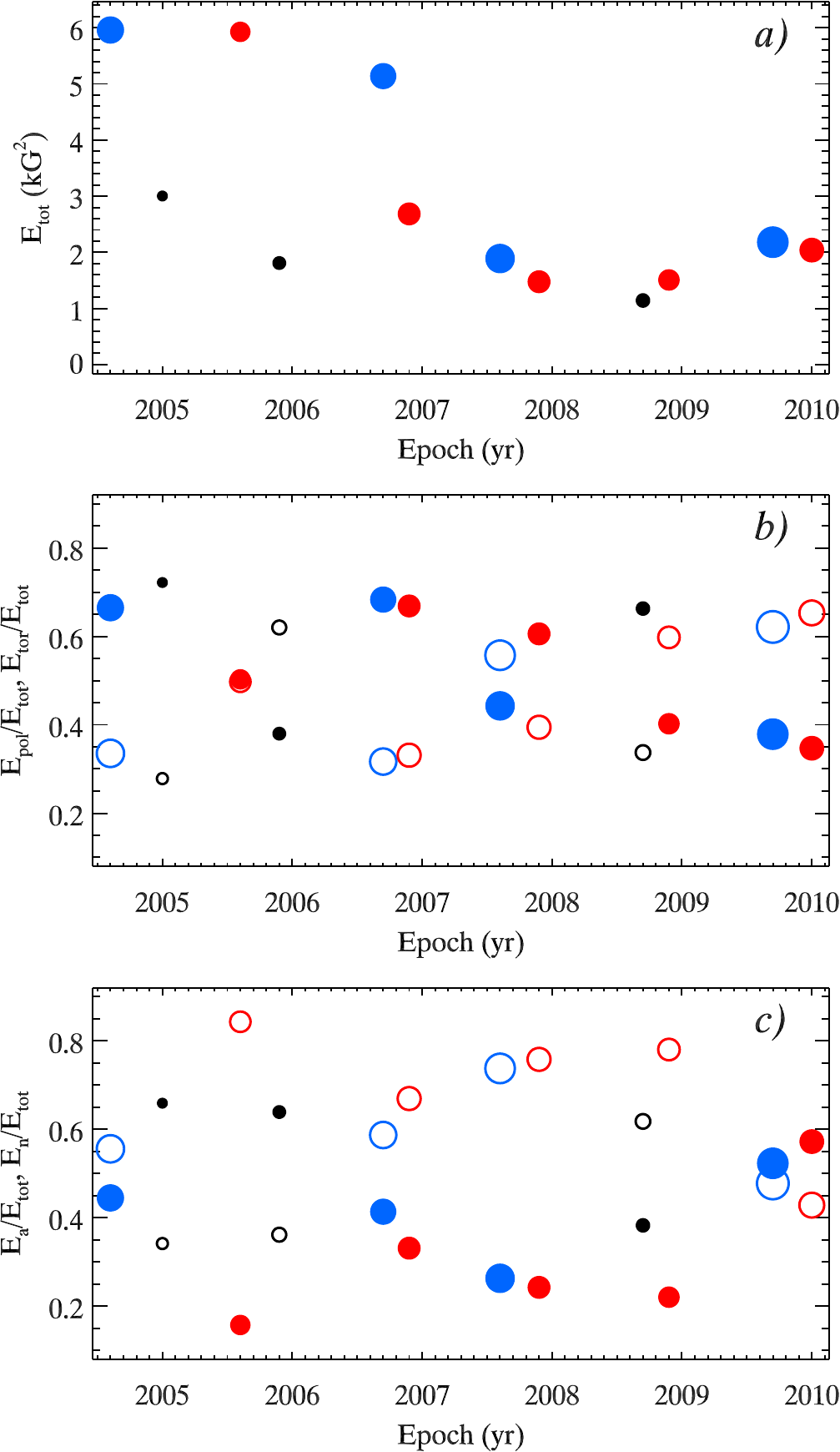}
\caption{Long-term evolution of different contributions to the magnetic field topology of \peg. The panels show: {\bf a)} variation of the total magnetic energy, {\bf b)} the relative contributions of the poloidal (\textit{filled symbols}) and toroidal (\textit{open symbols}) harmonic components, and {\bf c)} the relative contributions of the axisymmetric (\textit{filled symbols}) and non-axisymmetric (\textit{open symbols}) components. The relative sizes of symbols and their colours correspond to the phase coverage of individual data sets, similar to Fig.~\ref{fig:bz-range}.
}
\label{fig:bharm}
\end{figure}

The sum of the spherical harmonic coefficients squared is proportional to the total energy contained in the stellar magnetic field. Figure~\ref{fig:bharm}a illustrates how this parameter has changed for \peg\ between 2004 and 2010. Consistently with the results discussed above, we find a noticeable decrease of the total field energy in the period between 2004 and 2008, with a possible reversal of this trend afterwards. One can note that the total magnetic energy recovered from the data sets containing only a few spectra (e.g. 2005.0, 2005.9) is systematically underestimated compared to the trend defined by other maps. It is reassuring that the analysis of these data sets does not result in spurious strong magnetic field features.

The time dependence of the relative contributions of the poloidal and toroidal field components is illustrated in Fig.~\ref{fig:bharm}b. It appears that our observations of \peg\ reveal a cyclic change of the field topology on the time scale of a few years. Ignoring magnetic maps corresponding to epochs with a partial phase-coverage, one can conclude that \peg\ exhibited a predominantly poloidal field before 2007.6 and a mainly toroidal field afterwards. However, the difference between the energies of the two components is never much larger than 20--30\%.

Finally, Fig.~\ref{fig:bharm}c assesses relative contribution of the axisymmetric and non-axisymmetric field components as a function of time. Here we define axisymmetric harmonic components as those with $m<\ell/2$ and non-axisymmetric ones as $m\ge\ell/2$ \citep[e.g.][]{fares:2009}. With this definition, non-axisymmetric field dominates all the time, except for the last two observing epochs.

\begin{figure}[!th]
\fifps{8cm}{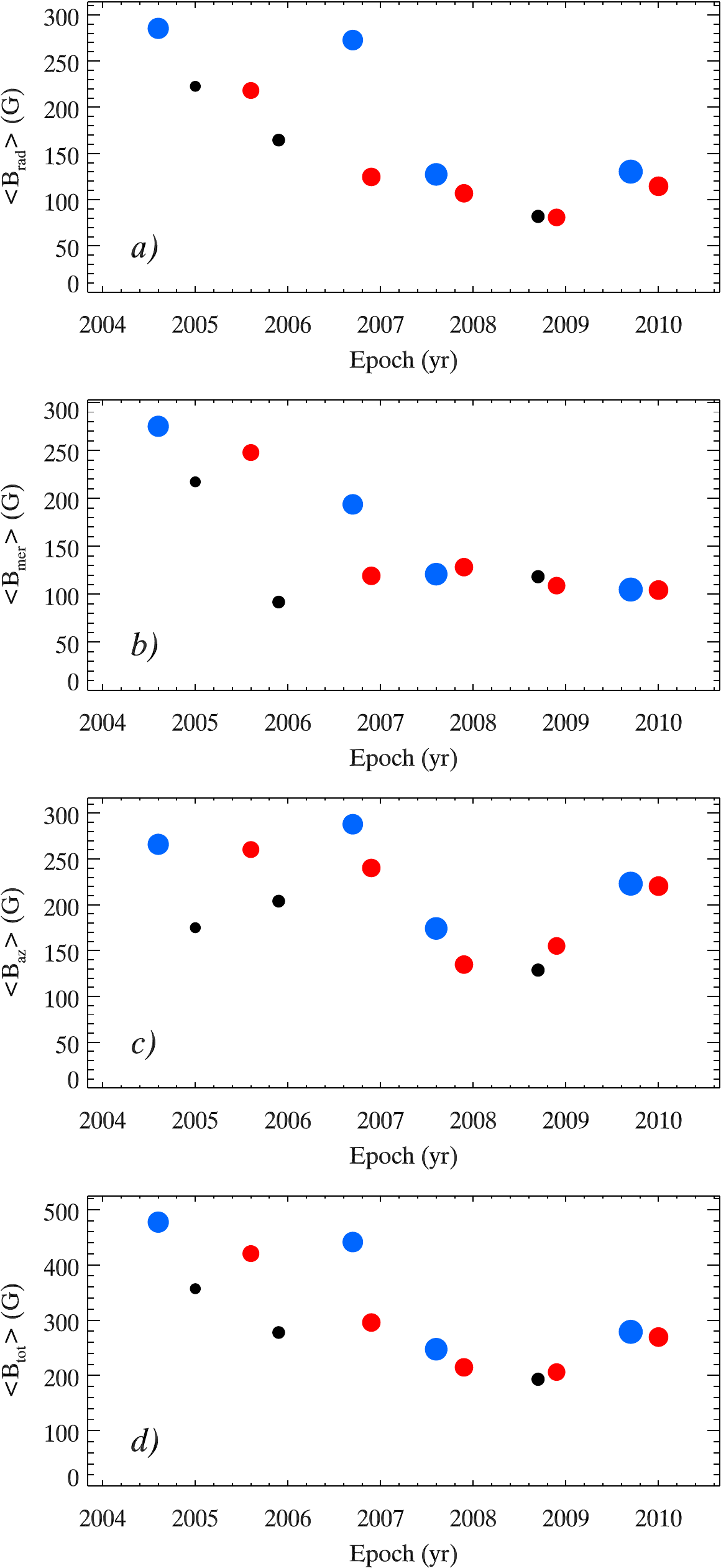}
\caption{Root-mean-square values of the different magnetic field components calculated over each ZDI map:
{\bf a)} radial field, {\bf b)} meridional field, {\bf c)} azimuthal field, {\bf d)} total field. 
The relative sizes of symbols and their colours correspond to the phase coverage of individual data sets, similar to Fig.~\ref{fig:bz-range}.
}
\label{fig:brms}
\end{figure}

As mentioned in Sect.~\ref{code}, formal error bars cannot provide a realistic estimate of uncertainties of the spherical harmonic coefficients recovered in a regularised least-squares problem. Instead, the scatter of points corresponding to close observational epochs gives an idea of the uncertainties. We can see that in Figs.~\ref{fig:brms} and \ref{fig:bharm} the points inferred from the poor phase-coverage data sets often deviate significantly from the general trends. On the other hand, there are only a few cases when results obtained from the good phase-coverage data exhibit abrupt changes. This suggests that the trends examined in this and the previous sections are real and are not dominated by random inversion errors.

\subsection{Extended magnetospheric structure}

The results of ZDI calculations are commonly used to investigate an extended stellar magnetospheric structure \citep{donati:2008b,jardine:2008,gregory:2008,fares:2012}. The knowledge of the field topology above the stellar surface and in the immediate circumstellar environment allows photospheric magnetic field measurements to be connected with the studies of stellar coronas, X-ray emission, and prominences and the interaction between the mass loss and magnetic field to be investigated. To determine the structure of the circumstellar magnetic field, one can use the potential field source surface (PFSS) extrapolation method developed for the solar magnetic field by \citet{van-ballegooijen:1998} and adapted for stellar magnetic fields by \citet{jardine:2002}. In this method the extended stellar magnetic field is represented as the gradient of a scalar potential expanded in a spherical harmonic series. The boundary condition for the radial field component at the stellar surface is provided by the empirical magnetic field maps. The outer boundary condition is given by a source surface of radius $R_{\rm s}$ beyond which the field is assumed to be purely radial.

We reconstructed the extended magnetospheric structure of \peg\ with the help of an independently developed PFSS code. This software was applied to all our ZDI maps. The source surface is placed at $R_{\rm s}=3 R_{\star}$, which is plausible given the mean value of the solar source surface radius of $2.5R_\odot$. Previous potential field extrapolation studies adopted similar $R_{\rm s}$ values for other cool active stars \citep[e.g.][]{jardine:2002,hussain:2002}.

\begin{figure*}[!th]
\centering
\fifps{17cm}{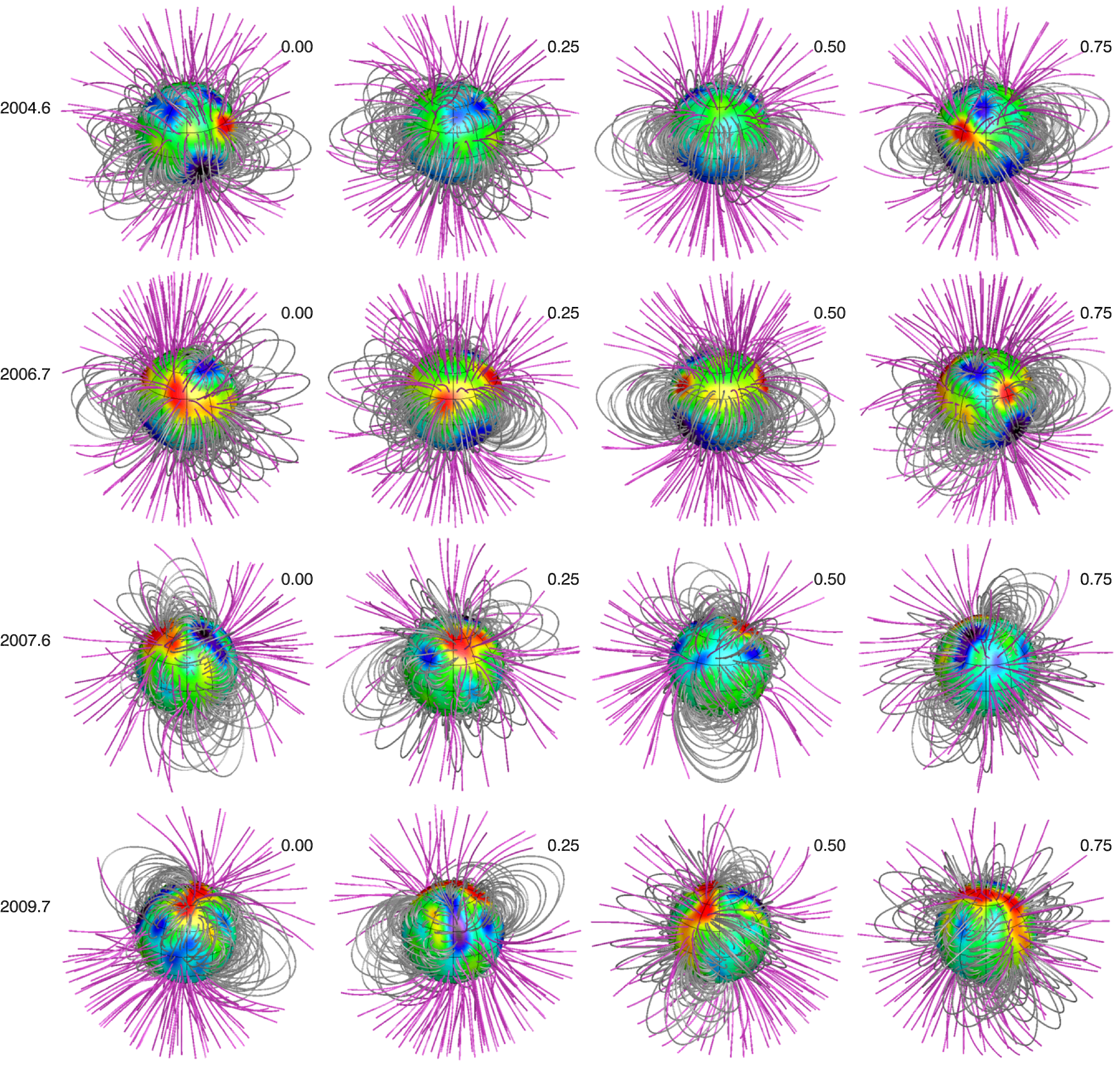}
\caption{Extended magnetosphere of \peg\ determined using potential field extrapolation from the radial component of ZDI maps. The star is shown at four equidistant rotation phases (\textit{columns}) for the four epochs with best phase-coverage (\textit{rows}). The underlying spherical map corresponds to the radial magnetic field component. The open and closed magnetic field lines are shown in different colour.
}
\label{fig:psse}
\end{figure*}

The magnetospheric structure of \peg\ predicted by our ZDI maps is illustrated in Fig.~\ref{fig:psse} for the four epochs with best phase-coverage. The stellar magnetosphere is shown at four distinct rotational phases, with different colours highlighting open and closed magnetic field lines. The evolution of the large-scale field is evident from this figure. Since the contribution of the small-scale (and hence less reliably reconstructed) field structures decays more rapidly with radius, this potential field extrapolation essentially presents a distilled view of the radial component of the ZDI maps, in which only the most robust large-scale information is retained. We find that during the first two epochs (2004.6 and 2006.7) the global field is nearly axisymmetric and is reminiscent of a dipole aligned with the stellar rotational axis. A drastic change of the large-scale magnetic topology occurs between epochs 2006.7 and 2007.6. Simultaneously with a sharp decline in the total magnetic field energy, the field becomes more complex and decidedly non-axisymmetric.

\section{Discussion}
\label{disc}

\subsection{Comparison with previous ZDI studies}

The only other RS\,CVn system repeatedly studied with ZDI is HR\,1099 (V711\,Tau). The papers by \citet{donati:1999b}, \citet{donati:2003}, and \citet{petit:2004a} presented magnetic field and brightness distributions recovered for about five epochs each, spanning the period from 1991 to 2002. In all these studies magnetic field reconstruction was carried out assuming an immaculate photosphere and the local line profiles were treated with a Gaussian approximation or using LSD profiles of slowly rotating inactive standards.

The observed LSD Stokes $V$ profiles of HR\,1099 have a typical peak-to-peak amplitude of 0.15\%, whereas the reconstructed field intensities are of the order of a few hundred G on average and reach up to $\sim$\,1~kG locally. This is comparable to our observational data and inversion results for \peg. The ZDI studies of HR\,1099 revealed dominant azimuthal magnetic fields, often arranged in unipolar rings encircling the star at a certain latitude. Repeated magnetic inversions suggested stability of these structures on the time scales of several years. The authors attributed these horizontal fields to a global toroidal magnetic component produced by a non-solar dynamo mechanism distributed throughout the stellar convection zone.

Compared to these studies of HR\,1099, our ZDI maps of \peg\ show a considerably smaller relative contribution of azimuthal fields. We still find a dominant toroidal component; however, these results are not fully equivalent nor easily comparable to those of, for example, \citet{donati:1999b} because here we use a harmonic representation of the magnetic field topology and hence are able to disentangle toroidal and poloidal contributions to the azimuthal field, whereas previous studies of HR\,1099 completely ignored the poloidal contribution to the azimuthal field. 

It is clear that our data contain no evidence of the persistent unipolar azimuthal ring-like magnetic structures similar to those reported for HR\,1099. Thus, either the dynamo mechanism operates differently in the two RS\,CVn stars with nearly identical fundamental parameters, or azimuthal fields may represent an artefact owing to a simplified ZDI approach adopted for HR\,1099 \citep{carroll:2009}. To this end, we note that any axially-symmetric structure appearing in stellar DI maps must be carefully examined to exclude possible systematic biases and, at least, must be connected to the stationary features in the observed profiles to prove its reality. Neither was  done for HR\,1099. Faced with this discrepancy between major features of the magnetic maps of \peg\ and HR\,1099, we conclude that a definite confirmation of the dominant ring-like azimuthal fields in the latter star and associated inferences about field generation mechanisms must await improved inversion methodologies and observations in all four Stokes parameters \citep{kochukhov:2011}.

\subsection{Relation between temperature and magnetic field maps}

Throughout the entire series of ZDI images reconstructed for \peg\ we see no obvious spatial correlation between the low-temperature spots and the strongest magnetic field features. This suggests that a significant fraction of magnetic flux is not associated with cool spots. Although a similar conclusion has been reached by previous studies \citep{donati:1997b,donati:1999b}, it was unclear whether this represented a genuine characteristic of an active-star stellar surface structure or an inversion artefact coming from an inconsistent modelling of the magnetic and temperature spots. Our work confirms the lack of the field-spot correlation based on a self-consistent and physically realistic analysis of the circular polarisation in spectral lines. Using a similar approach, \citet{carroll:2007} also failed to detect a strong correlation between magnetic and temperature features on the surface of \peg. These results may be interpreted as evidence that current ZDI maps are mostly sensitive to magnetic fields at photospheric temperature and entirely miss the very strong fields inside cool spots. Instead, numerical experiments predict that self-consistent ZDI should be capable of recovering fields inside cool spots even if the temperature contrast is as large as 1500~K \citep{kochukhov:2009c,rosen:2012}. 

A limited spatial resolution of the ZDI maps may be another reason for not seeing a link between fields and cool spots. Both the magnetic and temperature distributions obtained using a spectral inversion technique reveal only the largest-scale structures, which are probably not monolithic but consist of many smaller spots with different geometries and field polarities. Then a local correlation between low-temperature spots and magnetic fields may be washed out in the current generation of ZDI maps. One can note that the presence of a substantial unresolved small-scale magnetic flux implies a much stronger average magnetic field strength than the one inferred from the ZDI analysis of circular polarisation. Highly inconsistent results of ZDI and Zeeman splitting studies of low-mass stars represent an example of this situation for a different dynamo regime \citep{reiners:2009}.

At the same time, it is not entirely obvious from a theoretical standpoint that a one-to-one relation between the cool spots and magnetic fields stemming from the solar paradigm can be universally extended to other types of cool active stars. There is an increasing amount of theoretical evidence \citep[e.g.][]{chan:2007,kapyla:2011,mantere:2011} of a pure hydrodynamical instability leading to the generation of large-scale vortices in the rapid rotation regime. These structures have so far been found only in local Cartesian simulations of turbulent convection, the sizes of the vortices always being very close to the box size, suggesting that these structures may have globally significant spatial extents. Depending on the rotation rate, either cool, cyclonic vortices for intermediate rotation, or warm anticyclonic vortices are excited, the temperature contrast being of the order of ten percent. This instability might contribute to the generation of magnetic fields independent from temperature structures in rapid rotators. So far, however, these structures have been detected neither in more realistic spherical geometry nor in the magnetohydrodynamic regime \citep[see e.g.][]{kapyla:2012}.

\subsection{Interpretation in terms of dynamo theory}

The picture arising from an extensive set of previous photometric and
spectroscopic observations of \peg\ suggests that the surface
magnetic field of this object concentrates on one or two active
longitudes, i.e. is highly non-axisymmetric, and that these active
longitudes evolve dynamically over time. During the epoch 1994--2002,
a persistent drift of the active longitude has been confirmed
\citep{berdyugina:1998b,berdyugina:1999,lindborg:2011}. It is not detectable during 2004--2010
\citep{hackman:2012} which is supported by the ZDI maps presented in this
paper. Furthermore, our new results give an indication that the strength
of the magnetic field has been monotonically decreasing, at least
during 2004--2009; it seems that a minimum was reached at about
2009, after which the magnetic field strength started
increasing again. The analysis of the energy contained in the poloidal
and toroidal components as function of time shows that while
in the beginning of the dataset the field was predominantly poloidal, the
portion of the toroidal field is increasing nearly linearly with time,
and is dominating at the end of the dataset. This is also reflected by
the increasing contribution of the non-axisymmetric component
exceeding the energy contained in the axisymmetric modes for
2009--2010. All these findings together hint towards a possible
minimum in the star's magnetic activity cycle, during which the
magnetic field tends to be more poloidal and axisymmetric, accompanied
with the signature of the non-axisymmetric drifting dynamo wave
getting too weak to be detectable. Unfortunately, the datasets,
especially the series of ZDI maps, are too short to make decisive
conclusions on the cyclic nature of the magnetic field. In any case it
seems evident that the magnetic field on global scale is far from
static.

How can this be understood in terms of dynamo theory? In the solar
case, the internal rotation and its non-uniformities are known from
helioseismic inversions, while in the case of other stars, photometric
period variations, interpreted as indirect proxies of stellar surface
differential rotation, are normally much smaller than the solar
value. Theoretically this is conceivable, as it has been predicted that the
faster the star rotates, the smaller the non-uniformities in its
rotation rate will be \citep[see e.g.][]{kuker:2005}. This means that in the rapidly rotating late-type
stars the operation of the dynamo should rely more strongly on the
collective inductive action of convective turbulence (called the $\alpha$
effect), and less on the non-uniformities of the rotational velocity
(called the $\Omega$ effect), the dynamo therefore being more of the
$\alpha^2$ type than the $\alpha \Omega$ solar dynamo. The simplest of
these systems \citep[see e.g.][]{krause:1980} excite dynamo modes that
are non-axisymmetric but show no oscillations, although drifts of the
magnetic structure, i.e. azimuthal dynamo waves, are typical. As more
and more observational evidence on dynamically changing magnetic
fields is being gathered, it has become evident that this simple picture
is not adequate. It has been suggested that either these objects have
more differential rotation than predicted by theory
\citep[e.g.][]{elstner:2005}, or that the mean-field transport
coefficients describing the convective turbulence are far too
simple. Indeed, oscillating dynamo solutions in the $\alpha^2$ regime
have been found with more complex profiles
\citep{baryshnikova:1987,mitra:2010,kapyla:2012}.
We also note that some observational evidence exists supporting
  the idea of stars showing a larger amount of differential
  rotation than actually predicted by the theoretical models, although
  the discrepancy between observations and theory appears to be quite
  small \citep{hall:1991,collier-cameron:2007}. Direct numerical
  simulations in spherical geometry also show results consistent with
  the quenching of relative differential rotation with increasing
  rotation rate \citep{kapyla:2011a}, although it is
  still challenging to relate these models to real stars. Therefore,
  we cannot completely rule out the existence of enough differential
  rotation in rapid rotators to be significant for the dynamo
  mechanism.

For our purposes, it is relevant to compare our
observational results with direct numerical simulations of turbulent
convection. Extensive parameter studies have been performed
particularly in Cartesian geometry \citep{kapyla:2012}, while such
studies in global spherical geometry remain challenging
\citep{miesch:2009}. One clear shortcoming of the local Cartesian
models is that the differential rotation cannot
self-consistently emerge as a result of the modelling but needs to be
imposed. The global spherical models can grasp this aspect, but only a
few are successful in reproducing the solar rotation profile
\citep[see][]{miesch:2006}. The Cartesian studies clearly indicate
that oscillatory $\alpha^2$ dynamos are quite natural in the
rotation-dominated regime. Typical solutions
\citep[e.g.][]{kapyla:2012} show radial and azimuthal components of
nearly equal strengths, with a rough $\pi/2$ phase separation in the
cycle. Inclusion of shear into such a system has two principal
effects. Firstly, the azimuthal component grows in strength versus the
radial component, and quite often the components are in anti-phase, the sign
changes and minima occur simultaneously. The ZDI data
indicates that the azimuthal field component is not completely
synchronised with the radial and meridional fields, i.e. they do not
seem to grow/decline simultaneously. This, again, is more
consistent with the $\alpha^2$ scenario than the $\alpha \Omega$
picture.

The dominating non-axisymmetric topology, azimuthal
dynamo waves, and even the time dependence of the magnetic fields in
rapid rotators can be quite readily understood in terms of dynamo
theory relying only on the $\alpha$ mechanism;
$\alpha \Omega$ dynamos, in contrast, produce latitudinal dynamo
waves (such as the solar butterfly diagram), mostly axisymmetric
fields, and oscillatory solutions are the preferentially excited
ones \citep[see e.g.][]{steenbeck:1969}.
The presence of a significant axisymmetric contribution, as was
found in the case of \peg\ in this study, is difficult to explain with
a pure $\alpha^2$ dynamo mechanism. To ``axisymmetrise'' some
part of the dynamo solution, some differential rotation will probably
be needed; therefore our ZDI results most reasonably point towards an
$\alpha^2 \Omega$ dynamo operating on the object under study.

\begin{acknowledgements}
Doppler imaging calculations presented in this paper were carried out at the supercomputer facility provided to the Uppsala Astronomical Observatory by the Knut and Alice Wallenberg Foundation and at the UPPMAX supercomputer center at Uppsala University. OK is a Royal Swedish Academy of Sciences Research Fellow supported by grants from the Knut and Alice Wallenberg Foundation and from the Swedish Research Council. TH was funded by the research programme ``Active Suns'' at the University of Helsinki.
\end{acknowledgements}


\end{document}